\documentclass[twocolumn]{aa}
\usepackage[varg]{txfonts}

\usepackage{float}
\usepackage{graphicx}
\usepackage{multirow}
\usepackage{pdflscape}
\usepackage{booktabs}
\usepackage{color,amsmath,amssymb}
\usepackage{dcolumn}
\usepackage{geometry}
\usepackage{pifont}
\usepackage{fleqn}
\usepackage{hyperref}
\usepackage{url}
\usepackage[T1]{fontenc}
\usepackage{caption}
\DeclareUnicodeCharacter{00A0}{~}
\usepackage{subcaption}
\usepackage{ae,aecompl}
\usepackage{pdflscape}
\usepackage[normalem]{ulem}
\usepackage{array}
\usepackage{epsf}
\usepackage{epstopdf}
\usepackage{amssymb}
\usepackage{longtable}
\usepackage{epsfig}
\usepackage{gensymb}
\usepackage{tabularx}
\usepackage{lscape} 
\usepackage{makecell}

\usepackage{changepage}

\usepackage{units}

\usepackage[varg]{txfonts}
\usepackage{verbatim}

\begin{document}
        
        \title{Aluminium oxide in the atmosphere of hot Jupiter WASP-43b}

        \titlerunning{Aluminium oxide in the atmosphere of Hot Jupiter WASP-43b}
        
        \subtitle{}     
        
        \author{Katy L. Chubb\inst{1}\thanks{email: k.l.chubb@sron.nl} \and Michiel Min\inst{1} \and Yui Kawashima\inst{1} \and Christiane Helling\inst{3} \and Ingo Waldmann\inst{2}}
        
        \institute{SRON Netherlands Institute for Space Research, Sorbonnelaan 2, 3584 CA, Utrecht, Netherlands
                \and Department of Physics and Astronomy, University College London, London, WC1E 6BT, UK \and Centre for Exoplanet Science, University of St Andrews, Nort Haugh, St Andrews, KY169SS, UK} 
        

        
        
        
        \abstract{

                We have conducted a re-analysis of publicly available Hubble Space Telescope Wide~Field~Camera~3 (HST WFC3) transmission data for the hot-Jupiter exoplanet WASP-43b, using the Bayesian retrieval package Tau-REx.
                We report evidence of AlO in transmission to a high level of statistical significance ($>$~5~$\sigma$ in comparison to a flat model, and 3.4~$\sigma$ in comparison to a model with H$_2$O only). We find no evidence of the presence of CO, CO$_2$,  or CH$_4$ based on  the available HST WFC3 data or on Spitzer IRAC data. We demonstrate that AlO is the molecule that fits the data to the highest level of confidence out of all molecules for which high-temperature opacity data currently exists in the infrared region covered by the HST WFC3 instrument, and that the subsequent inclusion of Spitzer IRAC data points in our retrieval further supports the presence of AlO. H$_2$O is the only other molecule we find to be statistically significant in this region. AlO is not expected from the equilibrium chemistry at the temperatures and pressures of the atmospheric layer that is being probed by the observed data. Its presence therefore implies direct evidence of some disequilibrium processes with links to atmospheric dynamics. Implications for future study using instruments such as the James Webb Space Telescope (JWST) are discussed, along with future opacity needs. 
                Comparisons are made with previous studies into WASP-43b. 
                
        }
        
        \maketitle

\section{Introduction} \label{sec:intro}

WASP-43b 
has been the subject of many scientific studies in recent years ~\citep[e.g.][]{18MeTsMa.wasp43b,18LoKr.exo,17GaMa.exo,17KoShTa.exo,17KeCo.wasp43b,15KaShFo.wasp43b}, largely because it is one of only a few exoplanets to have observed emission phase curve data~\citep{14StDeLi.wasp43b} with strong evidence of molecular signatures, as demonstrated by \cite{14KrBeDe.wasp43b}. The planet is assumed to be tidally locked, which means that some information on atmospheric variability across the planet's surface can be gained via analysis of this emission data at different phases of planetary transit, making it a strong candidate for detailed studies of atmospheric circulation models. 
It
was discovered   by \cite{11HeAnCo.wasp43b}  around an active K7V star, with deduced planetary parameters from radial velocity measurements and transit observations of 2.034~$\pm$~0.052~$M_J$ and 1.036~$\pm$~0.019~$R_J$~\citep{12GiTrFo.wasp43b}. The presence of strong equatorial jets has been suggested by previous studies, such as \cite{15KaShFo.wasp43b}, largely due to the  strong day--night temperature contrast~\citep{14StDeLi.wasp43b,15KaShFo.wasp43b,17GaMa.exo,19IrPaTa.wasp43b} to explain the eastward hotspot shift of 12.3~$\pm$~1$\degree$ (corresponding to 40 minutes before the eclipse) observed by~\cite{14StDeLi.wasp43b}. 

The publicly available transmission and emission data for WASP-43b is primarily a result of observations by the Wide Field Camera 3 (WFC3) instrument on the Hubble Space Telescope (HST). WFC3 was used to observe three full-orbit phase curves, three primary transits, and two secondary eclipses (proposal ID 13467, PI: Jacob Bean~\citep{13Bean.hst}). The first light-curve fitting of the transmission data was carried out by \cite{14KrBeDe.wasp43b}, and later independently by \cite{18TsWaZi.exo}. We consider both sets of data in this work. Spitzer IRAC data measured at 3.6~$\mu$m and 4.5~$\mu$m is available from \cite{14BlHaMa.wasp43b}, with an independent re-analysis of the transit depth from \cite{19MoDaDi.wasp43b}. 
Previous analyses of the WFC3 data by \cite{14KrBeDe.wasp43b} and \cite{17StLiBe.wasp43b} have found evidence of H$_2$O in both transmission and emission, with \cite{17StLiBe.wasp43b}  deducing the presence of CO and/or CO$_2$ based on the Spitzer data points. \cite{19WeLoMe.wasp43b}  also find  evidence of H$_2$O in transmission. CH$_4$ was found to vary in emission with phase by \cite{17StLiBe.wasp43b}, with some caution on the derived abundances demonstrated by \cite{16FeLiFo.exo}.

AlO has been detected in oxygen-rich stars~\citep[e.g.][]{17BeDeRa.AlO,17TaKaTa.AlO}. 
There has, however, only been one previous observed indication of AlO in an exoplanet atmosphere, by \cite{19EsMaWe.wasp33b} in the atmosphere of the highly irradiated `super-hot Jupiter' WASP-33b.
They speculate about the presence of a thermal inversion; \cite{19GaMa.wasp43b} have also recently proposed AlO as a species that could cause a thermal inversion. As well as giving insight into the atmosphere, detecting heavy elements such as Al in a planetary atmosphere also gives some insight into planet formation processes~\citep{12JoLi.exo,14HaHi.exo}.

Although not explicitly stated, it is assumed that the study of \cite{19EsMaWe.wasp33b} uses the line list for AlO by \cite{15PaYuTe.AlO}, which was computed in 2015 as part of the ExoMol project~\citep{ExoMol_gen} and remains the only high-temperature line list for AlO suitable for retrievals of this kind.     
The line list is valid up to 8000~K, includes various electronic states~\citep{14PaHiTe.AlO}, covers the region 0.28~-~100~$\mu$m, and consists of over 5~million transitions, making it highly suitable for the characterisation of exoplanet or stellar atmospheres. The studies of \cite{17BeDeRa.AlO} and \cite{17TaKaTa.AlO}, on the other hand, rely on a handful of individual lines from rotational transitions and, although the source of their opacity data is again not explicitly stated, it is assumed that experimental data was used. 

This paper is structured as follows. In Section~\ref{sec:methods}, we provide details of our retrieval process and the statistical measures used, followed by the results in Section~\ref{sec:results}. We discuss various aspects of the results in Section~\ref{sec:discuss}, including the presence of clouds in Section~\ref{sec:clouds}, and equilibrium chemistry in Section~\ref{sec:eqchem}. In Section~\ref{sec:spitzer} we discuss the effects of including available transmission spectra from other instruments in different wavelength regions, which is followed by a comment on the currently available opacity data in Section~\ref{sec:opacity}. Our summary is given in Section~\ref{sec:summary}.


\section{Methods}\label{sec:methods}

\subsection{Transmission retrieval}\label{sec:trans_ret}

For the retrievals presented in this work, we use the Bayesian retrieval package Tau-REx~\citep{15WaTiRo.taurex}. Some preliminary tests and checks were conducted independently using Bayesian retrieval package ARCiS~\citep{19MiOrCh.arcis}. Both codes use the {\sc Multinest} \citep{08FeHo.multi,09FeGaHo.multi,13FeHoCa.multi} algorithm to sample the specified parameter space for the region of maximum likelihood. Of these two codes, only Tau-REx is currently publicly available. Full details on the ARCiS code are presented in a separate paper~\citep{19MiOrCh.arcis}. The most important information can be found in \cite{18OrMi.arcis}. The code consists of a forward modelling part based on correlated-k molecular opacities and cloud opacities using Mie and  distribution of hollow spheres (DHS; see \citealt{05MiHoKo.arcis}) computations. With ARCiS it is possible to compute cloud formation~\citep{18OrMi.arcis} and chemistry \citep{18WoHeHu.exo} from physical and chemical principles. The code has been benchmarked against petitCODE \citep{15MoBoDu.petitCODE} in \cite{18OrMi.arcis}. For the retrieval part the {\sc Multinest} algorithm is employed. Benchmarks for the retrieval have been performed in the framework of the ARIEL mission~\citep{18PaBeBa.ARIEL} showing excellent agreement with multiple other retrieval codes.

We have recently converted all molecular line list data available from the ExoMol~\citep{ExoMol_gen} and HITEMP~\citep{HITEMP} databases into cross sections and k-tables, for input into both Tau-REx and ARCiS. Cross sections and k-tables were computed at R=$\frac{\lambda}{\Delta\lambda}$=10,000 and R=300, respectively, on a grid of 27 temperatures between 100 and 3400~K, and 22 pressures between 1$\times$10$^{-5}$ and 100~bar. Details of the parameters and file formats used for these opacity data, which were converted into cross sections using ExoCross~\citep{ExoCross}, are  outlined in~\cite{19Chubbetal}, along with the publicly available opacities. 

In order to  fully assess which of these molecules is most likely to be causing the absorption features observed in the transmission spectrum of WASP-43b, we carried out the following steps, going from simple to more complex retrieval procedures: 
\begin{enumerate}

\item We first carried out a set of simple free retrievals that each include only one molecule, in order to subsequently exclude those with no absorption features in the WFC3 HST wavelength region (1.1~-~1.7~$\mu$m). 

\item Forward models for individual species, computed using ARCiS, are given in the Appendix (see Figure~\ref{fig:ARCiS_assess}), plotted alongside the transmission data for WASP-43b from \cite{14KrBeDe.wasp43b}. These were used in order to help assess which molecules to include in subsequent retrievals. These figures are intended to give an indication of where absorption features would occur in the HST WFC3 region for each of these species, and are not the results of the free retrievals  specified in step 1. 

\item We then assessed the reduced $\chi^2$ value for another set of simple retrievals, which each include only two molecules: H$_2$O plus one other molecule. For this we consider all the molecules that were found to exhibit some absorption features in the  WFC3 HST wavelength region, as determined in steps 1 and 2. ARCiS was used for steps 1~--~3.

\item We set up more complex retrievals using Tau-REx, the results of which are presented in Section~\ref{sec:results}.
Much of the set-up for these retrievals are as described in \cite{18TsWaZi.exo}, with a summary of the free and fixed parameters used in the present work given in Tables~\ref{t:free_pars} and \ref{t:fixed_pars}, respectively. We used free retrievals here with regard to the molecular abundances; i.e. no chemistry was assumed, and the volume mixing ratio for each molecular or atomic species was allowed to vary within the bounds specified by Table~\ref{t:free_pars}. 

\end{enumerate}

        
\begin{table*}[t]
        \caption{Free parameters used in the TauREx retrievals. The cloudy retrievals use the same parameters as the cloud-free ones, with the additions mentioned below.  }
        \label{t:free_pars} 
        \centering  
        \begin{tabular}{l|l|l|l}
                \hline\hline
                \rule{0pt}{3ex}Approach &       Parameter       &       Prior   & Description     \\
                \hline
\rule{0pt}{3ex}Cloud free  & log(molecule)  & -12 \dots 0 &  Molecular abundances \\
& $T_{\rm iso}$ ($K$) & 100 \dots 1800 & Isothermal temperature \\
& $R_{p}$ ($R_J$) & 0.017 \dots 1.055 &  Planetary radius at 10 bars \\
\hline
\rule{0pt}{3ex}Cloudy  & log(P$_{\rm top}$ (Pa))  & -3 \dots 6 &  Cloud top pressure \\
                \hline\hline
        \end{tabular}
\end{table*}

\begin{table*}[t]
                \caption{Fixed parameters used in the TauREx retrievals.   }
        \label{t:fixed_pars} 
        \centering  
        \begin{tabular}{l|l|l}
                \hline\hline
                \rule{0pt}{3ex}Parameter        &       Value   & Description   \\
                \hline
        \rule{0pt}{3ex}$T_{*}$ ($K$) & 4520  &   Stellar temperature$^1$ \\
        $R_{*}$ ($R_\odot$) & 0.667  &  Stellar radius$^1$ \\
        $M_{p}$ ($M_J$) & 2.034  &  Planetary mass$^1$  \\
        $M_{*}$ ($M_\odot$) & 0.717  &  Stellar mass$^1$ \\
        H$_2$ / He  & 0.17 &  (H$_2$ / He) ratio\\
        n$_{P_{\rm layers}}$ & 100  & Number of pressure layers  \\
        log(P$_{\rm layers}$ (Pa))& -5 \dots +6 & Range of pressure layers  \\
        CIA (H$_2$-H$_2$), (H$_2$-He) & HITRAN &  Collision induced absorption$^2$  \\
        \hline \hline
\end{tabular}
        {\flushleft
                $^1$ \cite{12GiTrFo.wasp43b};
                $^2$ \cite{HITRAN_2016,01BoJo.cia}
        }
\end{table*}

\subsection{Statistical measures}\label{sec:stat_measures}

In order to assess which molecule, or combination of molecules, is most likely to be causing the absorption features observed in WASP-43b, we use the following statistical measures. 

For step 3 of Section~\ref{sec:trans_ret}, the reduced $\chi^2$ value is used as part of the assessment to determine which molecules to include in subsequent retrievals. The reduced $\chi^2$ is a simple metric used to determine how well a particular model (in this case the results of our retrieval) fits a set of observed data. The use of reduced $\chi^2$, as opposed to $\chi^2$, means that retrievals using different number of molecular absorbers can be directly compared. The data we use here is the transmission spectra of WASP-43b from the HST WFC3 instrument, as analysed and presented in \cite{14KrBeDe.wasp43b}. A smaller reduced $\chi^2$ generally indicates a retrieval result that fits the observed data better, with a value~$<$~1 usually being an indication of over-fitting. Formally, models with a reduced $\chi^2$ closest to 1 are favoured over other models. However, we show through further assessments that this is not the case for, for example, the C$_2$H$_2$ + H$_2$O model. This model has a reduced $\chi^2$ close to 1, but the inclusion of C$_2$H$_2$  is found not to be significant when considering the Bayes factors of various models (see  discussion above and Section~\ref{sec:results}). We conclude that the use of reduced $\chi^2$ as a guide is limited and prone to error, and therefore a more rigorous approach is required. For this reason, we conduct the following Bayesian analysis for the full set of molecules used for this reduced $\chi^2$ assessment (see Section~\ref{sec:results} and Tables~\ref{t:sources} and \ref{t:bayes_factors_appendix}).


For step 4 of Section~\ref{sec:trans_ret}, we use a more rigorous way to determine the likelihood of a retrieval in comparison to the prior  base set-up: the Nested Sampling Global Log-Evidence (log($E$)). This is given as an output from the {\rm Multinest} algorithm  \citep{08FeHo.multi,09FeGaHo.multi,13FeHoCa.multi}. 
This Bayesian log-evidence is then used to find the Bayes Factor ($B_{01}$) \citep[see e.g.][]{15WaTiRo.taurex}, which is a measure to assess the significance of one model against another (here `model' refers to the set of free parameters used, in particular which molecular absorbers and whether clouds are included). If the natural log of the Bayes Factor, ln$(B_{01})$~$>$~5 then, according to \cite{08Trotta.stats}, the model can be considered significant with respect to the base  model; ln$(B_{01})$~$>$~5 corresponds to $>$~3.6~$\sigma$ detection over the base model, while ln$(B_{01})$~$>$~11 corresponds to $>$~5~$\sigma$ detection over the base model.

\section{Retrieval results}\label{sec:results}

The reduced $\chi^2$ values found in  step 2 of Section~\ref{sec:trans_ret} are given in Table~\ref{t:sources}, along with the line list data used for each molecule. AlO and H$_2$O were among the molecules with the smallest reduced~$\chi^2$.  It should be noted that the value of reduced $\chi^2$ itself is not exact and is prone to error, and so is only used here as a guide  to which molecules to include in subsequent retrievals. The models with $\chi^2<1$ (usually an indicator of over-fitting) cannot be distinguished from one another.

\begin{table*}[t]
        \caption{Reduced $\chi^2$ for different combinations of molecules (in addition to H$_2$O) included in a cloud-free retrieval using ARCiS, and references for the line lists used. Retrievals with only AlO or only H$_2$O are also shown, for comparison.   }
        \label{t:sources} 
        \centering  
        \begin{tabular}{l|l|l}
                \hline\hline
                \rule{0pt}{3ex}Molecule (in addition to H$_2$O) &       Reduced $\chi^2$        &       Line list data used     \\
                \hline
\rule{0pt}{3ex}AlO & 0.8 &  \\
AlO + Na & 0.83 &  \\
AlO + Na + CH4 & 0.88  \\
AlO (only) & 0.98 & \cite{14PaHiTe.AlO} \\
C$_2$H$_2$ & 1.04 & \cite{ExoMol_C2H2} \\
TiO & 1.07 & \cite{ExoMol_TiO} \\
FeH & 1.07 & \cite{10WeReSe.FeH}\\
H$_2$O (only) & 1.09 & \cite{ExoMol_H2O}\\
K & 1.11 & \cite{NISTWebsite} \\
Na & 1.12 & \cite{NISTWebsite} \\
HCN & 1.14 & \cite{ExoMol_HCN}\\
HeH+ & 1.14 & \cite{19AmDiJo.HeHplus}\\
CH$_4$ & 1.15 & \cite{ExoMol_CH4}\\
CO$_2$ & 1.15 & \cite{HITEMP} \\
C$_2$H$_4$ & 1.15 & \cite{18MaYaTe.C2H4}\\
NH$_3$ & 1.15 & \cite{jt771}\\
CH & 1.15 & \cite{14MaPlVa.CH}\\
H$_2$CO & 1.15 & \cite{15AlYaTe.H2CO}\\
H$_2$S & 1.15 & \cite{ExoMol_H2S}\\
OH & 1.16 & \cite{18YoBeHo.OH} \\
HNO$_3$ & 1.16  & \cite{15PaYuTe.HNO3}\\
TiH & 1.19 & \cite{05BuDuBa.TiH} \\
ScH & 1.2 & \cite{15LoYuTe.ScH} \\
VO & 1.23 & \cite{ExoMol_VO}\\
MgO & 1.25 & \cite{19LiTeYu.MgO} \\
No mols & 1.63 \\
                \hline\hline
        \end{tabular}
\end{table*}

The full retrievals that were outlined in step 3 of Section~\ref{sec:trans_ret}, were performed using Tau-REx (see Tables~\ref{t:free_pars} and~\ref{t:fixed_pars} for the free and fixed parameters used, respectively). 
Table~\ref{t:bayes_factors} gives a  summary of the Nested Sampling Global Log-Evidence (see Section~\ref{sec:stat_measures}) of various retrievals, along with the natural log of the Bayes factor, ln$(B_{01})$, and $\sigma$ likelihood against: a flat base retrieval (i.e. one with no molecular features), a retrieval with only H$_2$O included, and a retrieval with only AlO included. Here the transmission spectra of WASP-43b from the HST WFC3 instrument is used, as analysed and presented in \cite{14KrBeDe.wasp43b}. 

 It can be seen that the best model against a flat spectra is that where both AlO and H$_2$O are included in the retrieval, which is preferred over a flat-line base model at over 5~$\sigma$. The findings of Table~\ref{t:bayes_factors} show that the presence of AlO in the model gives more of a statistical improvement to the fit than the inclusion of H$_2$O; a model with both H$_2$O and AlO is preferred over a model with only H$_2$O  at a confidence level of 3.4~$\sigma$, whereas a model with H$_2$O and AlO is preferred over a model with only AlO at a confidence level of 2.6~$\sigma$. Table~\ref{t:bayes_factors_appendix} in the Appendix gives ln$(B_{01})$ for H$_2$O + each molecule which is considered in Table~\ref{t:sources}. The line list sources are  given in Table~\ref{t:sources}. In all these models we include Rayleigh scattering and collision induced absorption (CIA) of H$_2$-He and H$_2$-H$_2$ \citep{HITRAN_2016,01BoJo.cia}. In order to check whether the inclusion of these continuum opacities has a significant effect on our results, we performed a series of retrievals with and without their inclusion. It can be seen from Table~\ref{t:bayes_factors_cia_R} that although there is some small variation in the Nested Sampling Global Log-Evidence for different combinations of including or not including CIA, Rayleigh scattering and clouds (for  models with H$_2$O only and with H$_2$O~+~AlO), we find that the inclusion of CIA and Rayleigh scattering does not significantly affect the results, and that the H$_2$O~+~AlO model is preferred over 
   the H$_2$O-only model for all combinations.
\begin{table*}[t]
        \caption{Nested Sampling Global Log-Evidence (log($E$))
                 of various retrievals of the HST/WFC3 data, along with the natural log of the Bayes factor,  ln$(B_{01})$, and $\sigma$ likelihood against: a flat `base' retrieval (i.e. with no molecular features), a retrieval with only H$_2$O included, and a retrieval with only AlO included. For reference, the Nested Sampling Global Log-Evidence of the base flat retrieval is 163.7, for the H$_2$O-only retrieval it is 173.2, and for the AlO-only retrieval it is 175.4. The values for $\sigma$ have been interpolated from Table~2 of \cite{08Trotta.stats}.}
        \label{t:bayes_factors} 
        \centering  
        \begin{tabular}{l | l | l | l | l}
                \hline\hline
                \rule{0pt}{3ex}Molecules included       &       log($E$)        & Clouds? & ln$(B_{01})$ &        $\sigma$ \\
                \hline
                \multicolumn{5}{l}{\rule{0pt}{3ex}Compared to flat model} \\
                \hline
                \rule{0pt}{3ex}H$_2$O &       173.2   & No  & 9.5 & 4.7         \\      
                H$_2$O &       172.9   & Yes & 9.2 & 4.6        \\      
                AlO    &       175.4  & No &  12.3 & $>$~5         \\      
                AlO    &       175.3   & Yes & 11.6 & $>$~5         \\      
                AlO + H$_2$O   &       177.4   & No &  13.7 & $>$~5              \\
                AlO + H$_2$O   &       176.7   & Yes & 13 & $>$~5            \\
                \hline
                \multicolumn{5}{l}{\rule{0pt}{3ex}Compared to H$_2$O-only model} \\
                \hline
                \rule{0pt}{3ex}AlO + H$_2$O    &       177.4   & No  & 4.2 & 3.4          \\      
                AlO + H$_2$O    &       176.7   & Yes & 3.8 & 3.2            \\      
                FeH + H$_2$O    &       174.1  & No &  0.9  & -              \\     
                TiO + H$_2$O    &       173.3 & No  & 0.1 & -               \\      
                C$_2$H$_2$ + H$_2$O     &   172.8    & No &  -0.4 & -       \\   
                VO + H$_2$O     &       172.1 & No &  -1.1 & -       \\     
                \hline
                \multicolumn{5}{l}{\rule{0pt}{3ex}Compared to AlO-only model} \\
                \hline
                \rule{0pt}{3ex}AlO + H$_2$O    &       177.4   & No  & 2.0 & 2.6       \\      
                AlO + H$_2$O    &       176.7   & Yes & 1.3 & 2.2            \\
                \hline \hline
        \end{tabular}
\end{table*}

\begin{figure}[]
        \includegraphics[width=0.45\textwidth]{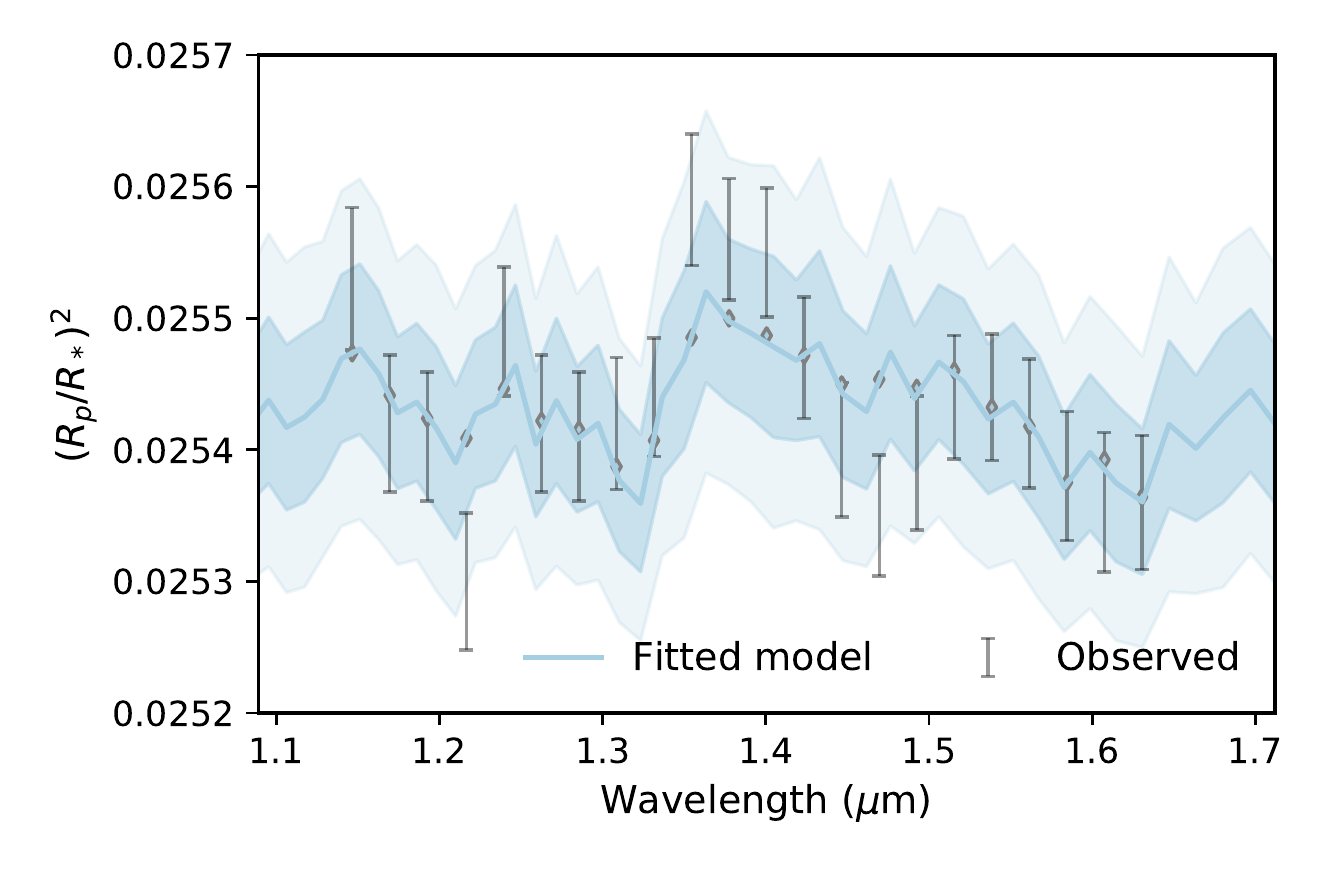}
        \includegraphics[width=0.45\textwidth]{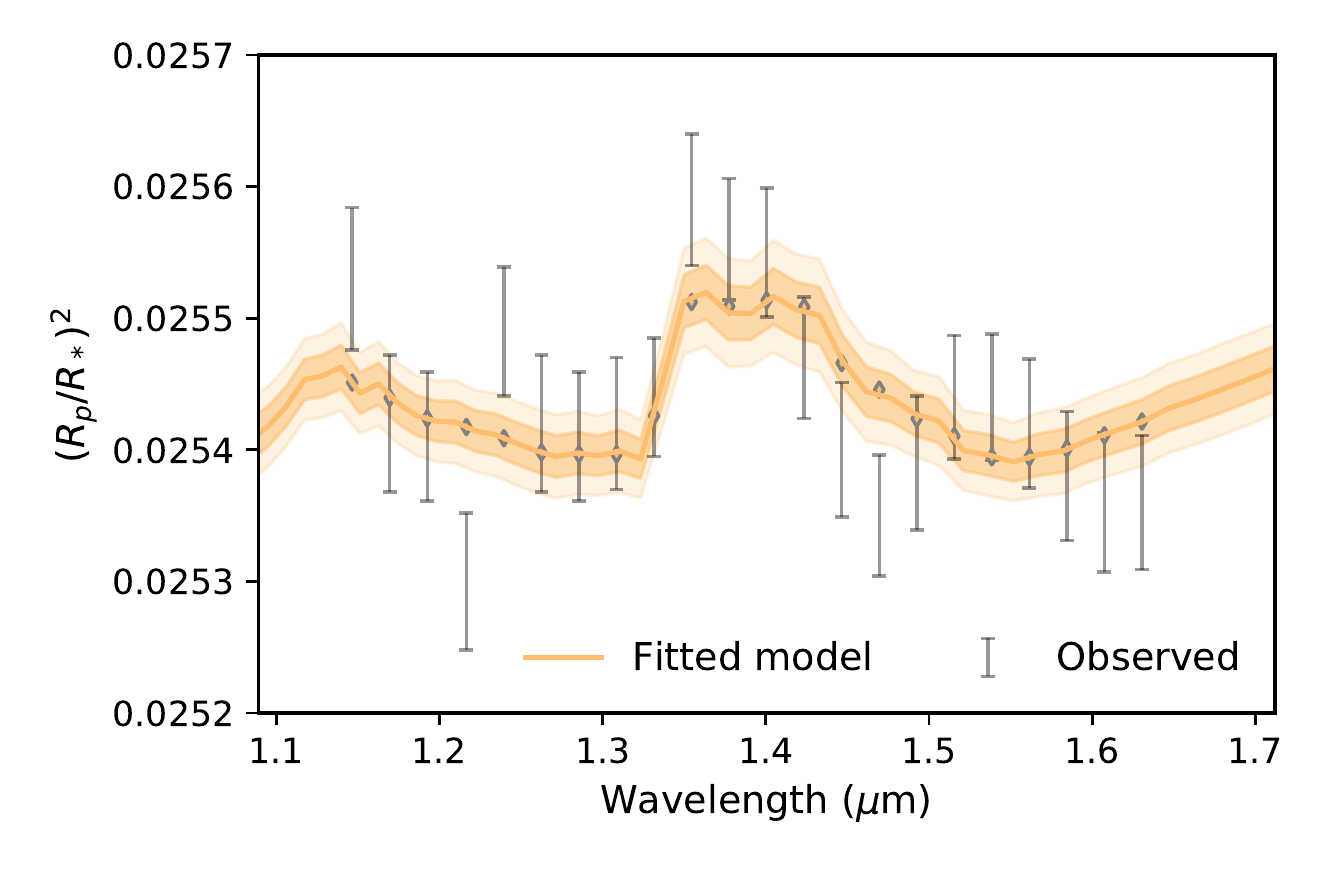}
        \caption{Cloud-free Tau-REx transmission retrieval results with H$_2$O and AlO (top) and with H$_2$O only (bottom). The different shading corresponds to 1~$\sigma$ and 2~$\sigma$ regions. }
        \label{fig:TauRex_trans_comp}
\end{figure}

\begin{figure}[]
        \includegraphics[width=\linewidth]{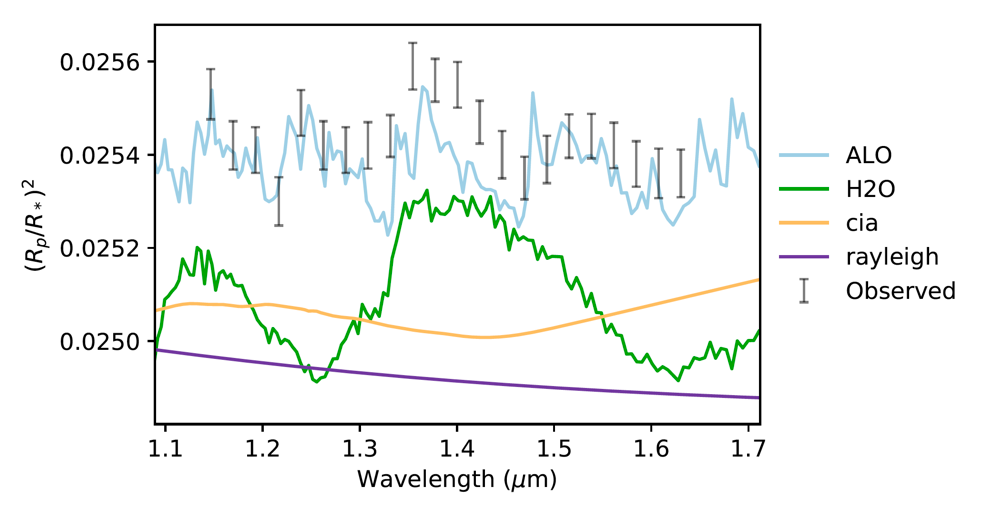}
        \centering
        \caption{Contributions of molecular features to the cloud-free Tau-REx retrieval results that include AlO and  H$_2$O only. }
        \label{fig:TauRex_h2o_alo_1m_contrib}
  \end{figure}

\begin{figure}[]
        \centering
        \includegraphics[width=\linewidth]{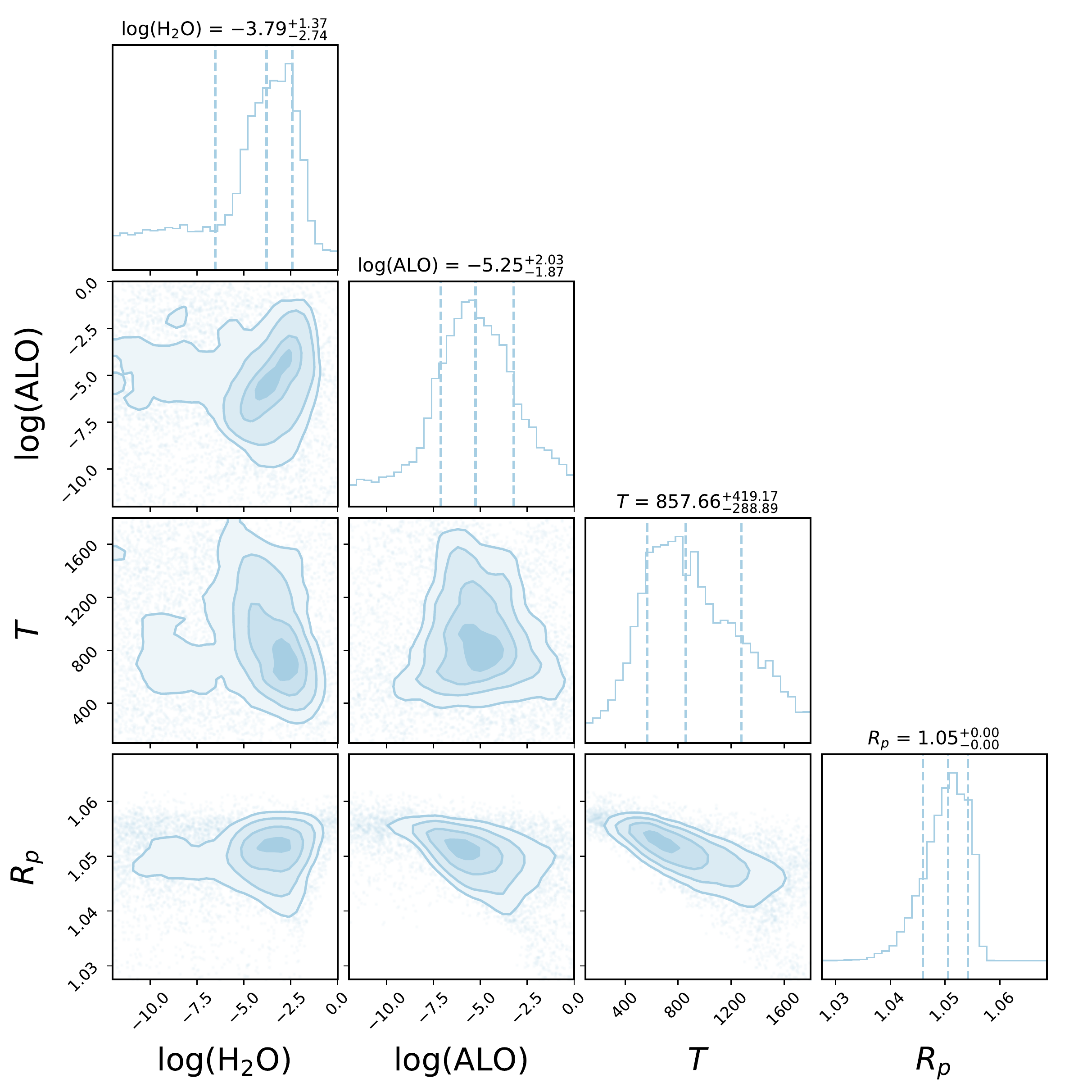}
        \caption{TauRex transmission retrieval posteriors for H2O and AlO, with no clouds.}
        \label{fig:TauRex_post1}
\end{figure}

Figure~\ref{fig:TauRex_trans_comp} shows the results of a retrieval which includes AlO and H$_2$O (top panel) and a retrieval only including H$_2$O (bottom panel). Figure~\ref{fig:TauRex_h2o_alo_1m_contrib} illustrates the contributions of molecular features in the former.
The posterior distributions of this cloud-free model with AlO and H$_2$O only are given in Figure~\ref{fig:TauRex_post1}. For comparison, we ran the same models using the outputs (based on the same observations) of the transit light-curve fitting by \cite{18TsWaZi.exo}. Although the Nested Sampling Global Log-Evidence was higher in all cases for the \cite{18TsWaZi.exo} data (most likely due to the higher number of derived data points), the Bayes factor was consistent with those presented in Table~\ref{t:bayes_factors}. We only include those models with the highest Bayes factors here; the inclusion of other molecules consistently gave negative (or <~1) Bayes factors in comparison to the H$_2$O + AlO model, indicating that their inclusion is not statistically favoured (see Table~\ref{t:bayes_factors_appendix} for the full list). The only exception here is that a model with H$_2$O + AlO + Na gave a weak-to-moderate detection of Na when using the high-res data from~\cite{18TsWaZi.exo} (a Bayes factor of 2.2, corresponding to $\sim$~2.6$\sigma$, in comparison to the same model with Na discluded). The same finding does not apply when using the data from \cite{14KrBeDe.wasp43b}; the far left data point on the former is not included in the latter, presumably due to concerns about the reliability of data from the edges of the WFC3 wavelength window. We therefore do not find any strong justification to include Na in our models. 
We also tried various retrievals with and without clouds, and found no strong evidence to justify including clouds in our model; the inclusion of clouds resulted in a negative Bayes factor, which demonstrates that the inclusion of extra parameters is not justified (based on the present data quality and number of observed data points) by a corresponding improvement in the fit.

\newpage

\section{Results}\label{sec:discuss}

No other molecules apart from H$_2$O and AlO were found to be statistically significant from our retrievals of the HST WFC3 data (or from the inclusion of Spitzer IRAC data) for WASP-43b. This is most likely either because  other molecular species that are present do not have strong absorption features in the region of the WFC3 data or because  they do but they are present in an abundance too low to be strongly detectable. It should be noted that the WFC3 transmission data is only probing a small layer in the upper atmosphere at the terminator regions of the planet, mainly in the region of 10$^{-3}$~-~10$^{-1}$~bar. 
In this section we discuss various aspects of our findings. 



\subsection{Clouds}\label{sec:clouds}
Although we did not find any statistical reason to include clouds in these retrievals, it is of interest to explore further the presence and type of clouds present on WASP-43b, as is done in~\cite{19HeGrSa.wasp43b}. It is possible that there are clouds, but they are either lower in the atmosphere, they do not cover the whole planetary surface, or they are so thin as to appear transparent. The models of \cite{16PaFoSh.exo} suggest that clouds are expected to always be present on the nightside of hot Jupiters, and previous studies such as \cite{18MeMaDe.wasp43b} and \cite{17KoShTa.wasp43b} indicate that there is a thick cloud layer on the nightside of WASP-43b. Works such as \cite{20VePaBl.wasp43b} point out the difficulty in differentiating between a cloudy and cloud-free model by retrieving HST/WFC3 data alone. Their models suggest the nightside of WASP-43b is cloudy, and the cloud coverage of the dayside depends on various microphysical processes and dynamics. 

 Cloud formation in giant gas planets occurs from a chemically very rich atmospheric gas causing the formation of cloud particles that are made of a whole mix of materials, including Mg/Si/Fe/O-solids and also high-temperature condensates like TiO$_2$[s] and Al$_2$O$_3$[s], e.g. in HD~189733~b and HD~209458~b~\citep{16HeLeDo.exo}.  
 The amount by which these cloud particles deplete the local element abundance depends on the  local thermodynamic properties of the atmosphere, which in turn determines the gas composition, and hence the abundance of gas species such as AlO. A variety of cloud species are thought likely to be present in the atmosphere of WASP-43b, including corundum, Al$_2$O$_3$[s]~\citep{19HeGrSa.wasp43b}. More detailed studies are required to determine in detail the interplay between local thermodynamics and cloud formation which would allow the presence of AlO in sufficient amounts to explain the findings of this work.

\subsection{Chemistry}\label{sec:eqchem}

There are a few aluminium hosting molecules that equilibrium chemistry models such as GGchem~\citep{18WoHeHu.exo} predict are more abundant than AlO at the temperatures and pressures expected in the region of the atmosphere being probed by the WFC3 transmission spectrum of WASP-43b, if solar elemental abundances are assumed. The most notable of these molecules are AlH, AlCl, AlF, Al$_2$O, AlOH, and atomic Al (which does not become Al+ until higher temperatures, around 3000~K). These would all ideally be included in further retrievals for the transmission spectra of WASP-43b. The availability of opacity data for these and other molecules will be discussed in Section~\ref{sec:opacity}. 

The presence of AlO in the region of the atmosphere we are probing (at the terminator regions, around 10$^{-3}$~-~10$^{-1}$~bar, i.e. relatively high in the atmosphere) is an indication of some disequilibrium processes at work in the atmosphere of WASP-43b; there has  been speculation about such processes in the atmospheres of hot-Jupiter exoplanets similar to WASP-43b \citep[see e.g.][]{10StHaNy.diseq}.  Although we do not know the exact cause of the disequilibrium processes at work in WASP-43b, the presence of AlO is most likely due to either vertical or horizontal mixing; equilibrium chemistry models predict AlO to be present at high abundance deeper in the atmosphere than the relatively high-up layers being probed by HST WFC3, and aluminium-bearing cloud species such as Al$_2$O$_3$[s] are expected to be present across the planet's atmosphere~\citep{19HeGrSa.wasp43b}. 
It is therefore either possible for turbulence to be dredging up gases towards the top of the atmosphere, and therefore causing the apparent deviation from equilibrium chemistry, or for the evaporation of clouds to be creating AlO gas in the hottest parts of the atmosphere, which could be horizontally transported to the terminator regions we are observing by the strong equatorial jets which have been suggested by previous studies, such as \cite{15KaShFo.wasp43b}. It has also been shown by \cite{14AgPaVe.atmos} that, for hot Jupiters similar to WASP-43b, horizontal mixing causes the volume mixing ratio of molecules at the terminator regions to become quenched towards values typical of the hottest dayside region.
More detailed studies, however, are required to determine in detail the interplay between local thermodynamics and cloud formation which would allow the presence of AlO in sufficient amounts to explain the findings of this work. Although, as mentioned above, the dominant Al-binding species in equilibrium is AlOH, relatively little is known about the kinetic and photochemistry of such metal binding species. It has been demonstrated the AlO plays a key role in forming clusters such as (Al$_2$O$_3$)$_n$ in AGB star outflows \citep{19BoGoDe.al}, and that AlO has been identified in the cold envelopes of AGB star R~Dor based on kinetic gas-phase simulations \citep{17DeRiWa.al}. It is therefore possible that AlO in an exoplanet atmosphere could be a indicator of kinetic chemistry, which affects metal-containing species. We thus plan to assess the potential effect of mixing for WASP-43b in future work, combined with more detailed cloud formation models, in order to assess the validity of various disequilibrium processes.


 

The asymmetry of the phase curve emission data for WASP-43b suggests some variation in molecular signatures across the planetary surface~\citep{17StLiBe.wasp43b}, as has been recently demonstrated for HAT-P-7b~\citep{19HeIrCo.exo}. It should be noted that this planet is considerably more irradiated than WASP-43b. More investigation is needed to determine whether AlO is observed in emission and in transmission, which, due to available emission phase spectroscopic data for WASP-43b, would give further insight into the varying abundances of different gases throughout different latitudes and altitudes of the planetary atmosphere. 


\subsection{Spitzer and other observational data}\label{sec:spitzer}


As  mentioned above, Spitzer data measured at 3.6$\mu$m and 4.5$\mu$m are available from \cite{14BlHaMa.wasp43b}, with an independent re-analysis of the transit depth from \cite{19MoDaDi.wasp43b}. 
We did not include these points in the initial retrieval due to large variation in their deduced transit depths depending on the method used to fit the data~\citep[see e.g.][]{19MoDaDi.wasp43b}. 
The dip at 3.6~microns and increased absorption at 4.5~microns was found by \cite{14KrBeDe.wasp43b} to be consistent with either CO or CO$_2$. It should be noted that this dip is less pronounced with the Spitzer data analysed by \cite{19MoDaDi.wasp43b}. Other molecules that exhibit a similar dip at 3.6~microns and increased absorption at 4.5~microns include SiO~\citep{ExoMol_SiO}, AlF~\citep{18YoBe.exo}, CaF~\citep{18HoBe.CaF}, LiCl~\citep{18BiBe.LiCl}, NS~\citep{18YuBoGo.NS},  PO, and PS~\citep{17LaPaLo.PO}. The current data availability for these molecules is  discussed in Section~\ref{sec:opacity}. 

\begin{figure}[]
        \includegraphics[width=0.43\textwidth]{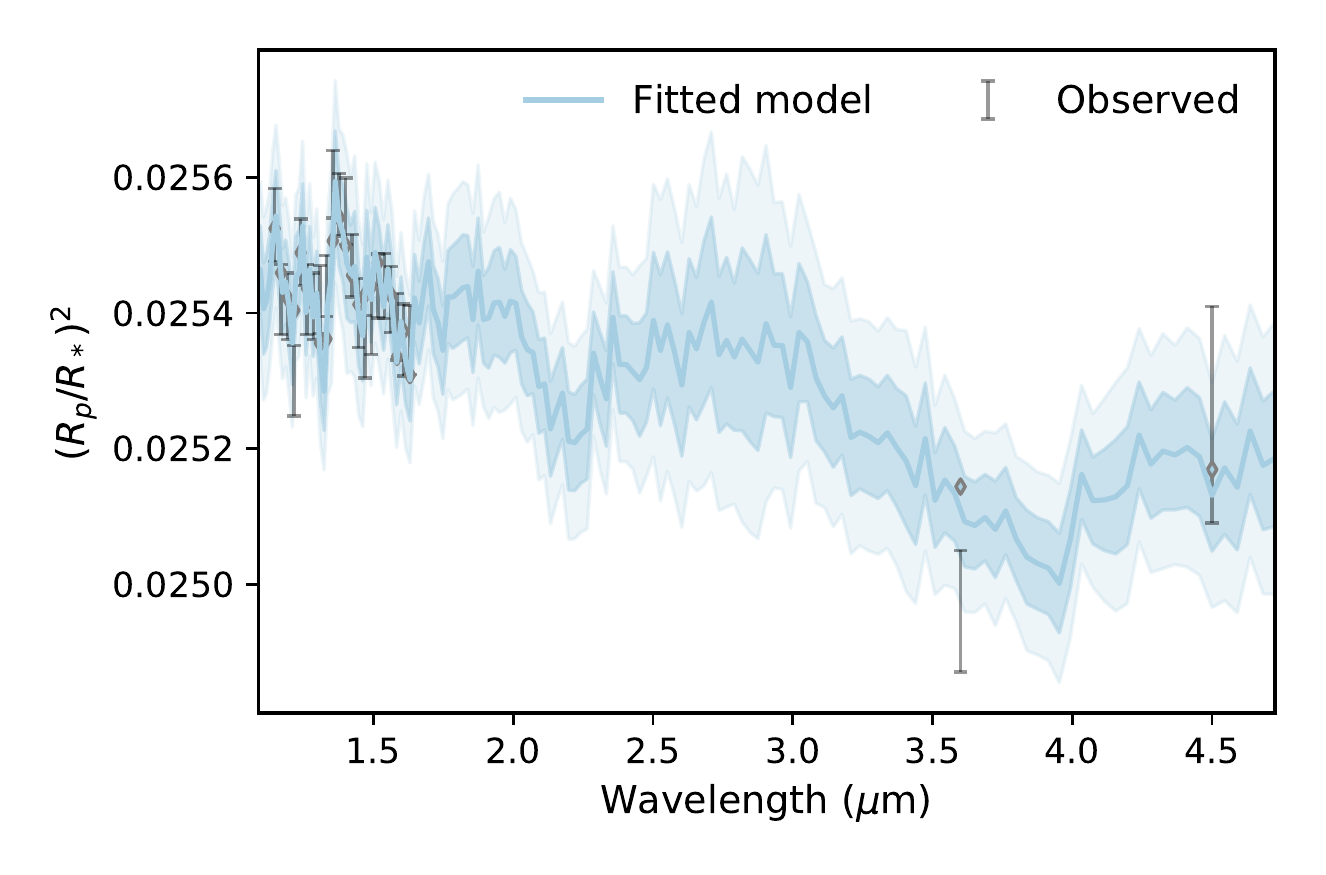}
        \includegraphics[width=0.57\textwidth]{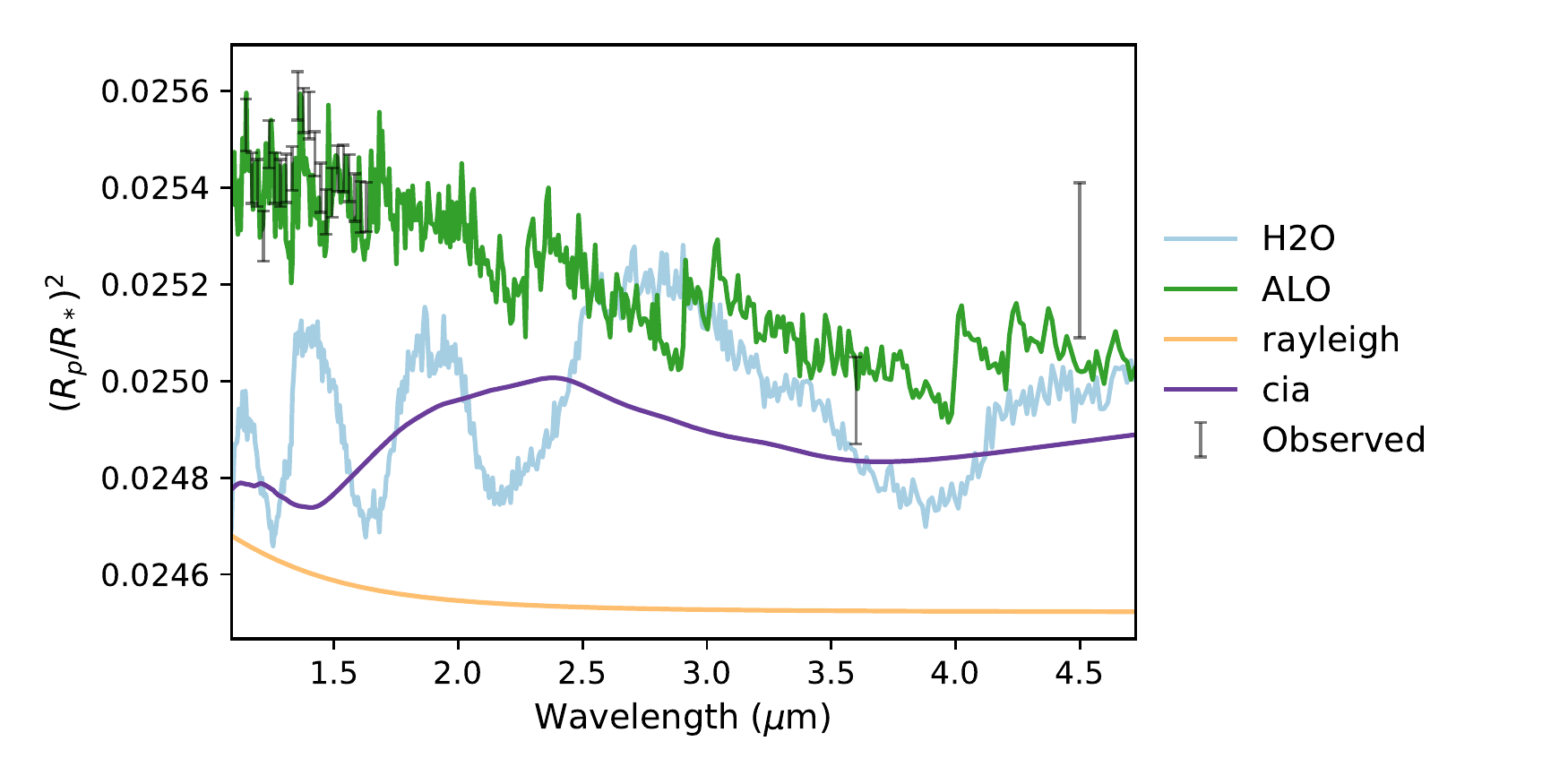}
        \caption{Tau-REx transmission retrieval results, including Spitzer points from \cite{14BlHaMa.wasp43b}, with H$_2$O and AlO (top) and the contributions of each molecule to the retrieval result (bottom). The different shading in the first panel corresponds to 1~$\sigma$ and 2~$\sigma$ regions. }
        \label{fig:TauRex_spitzer}
\end{figure}

\begin{figure}[]
        \includegraphics[width=0.43\textwidth]{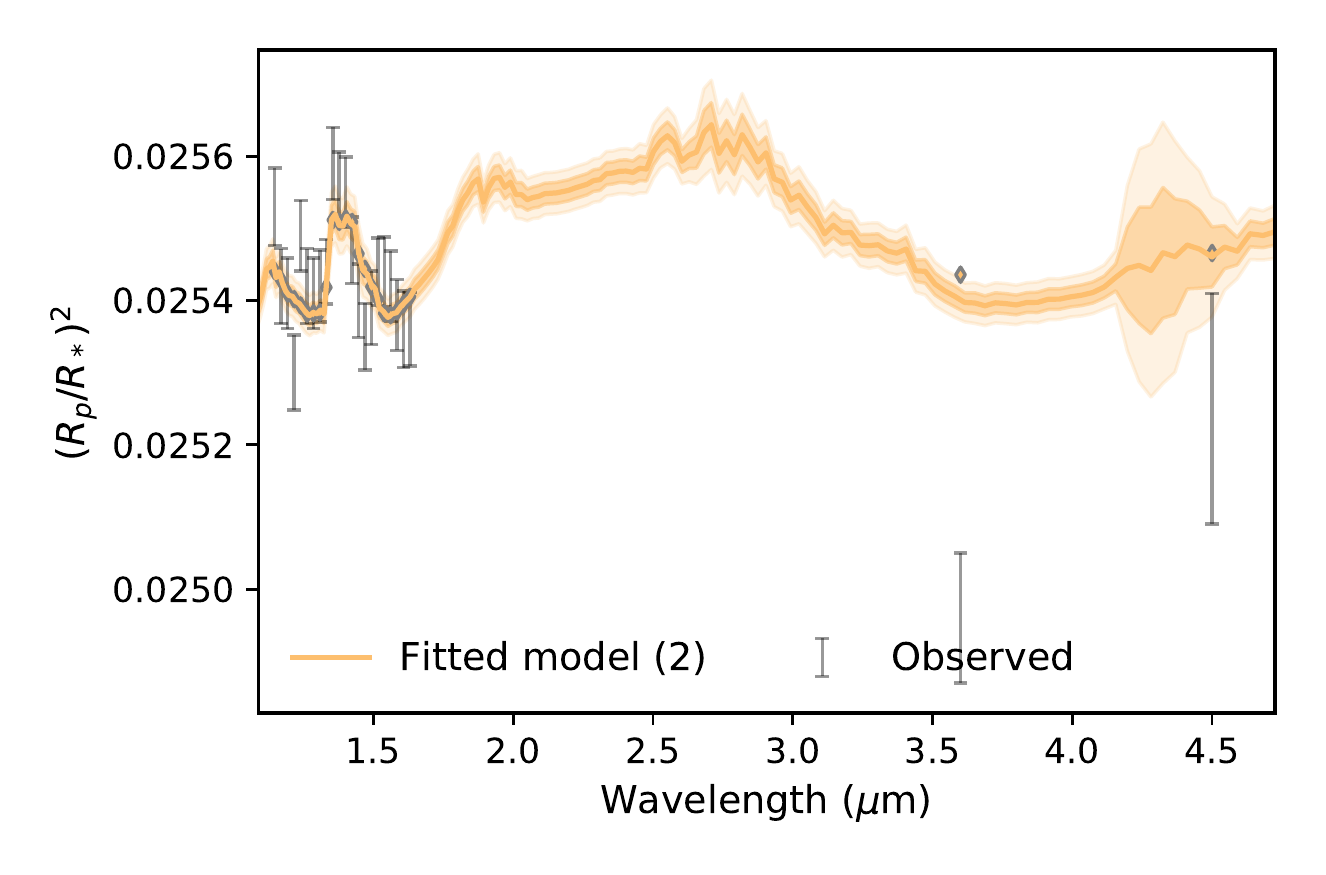}
        \includegraphics[width=0.57\textwidth]{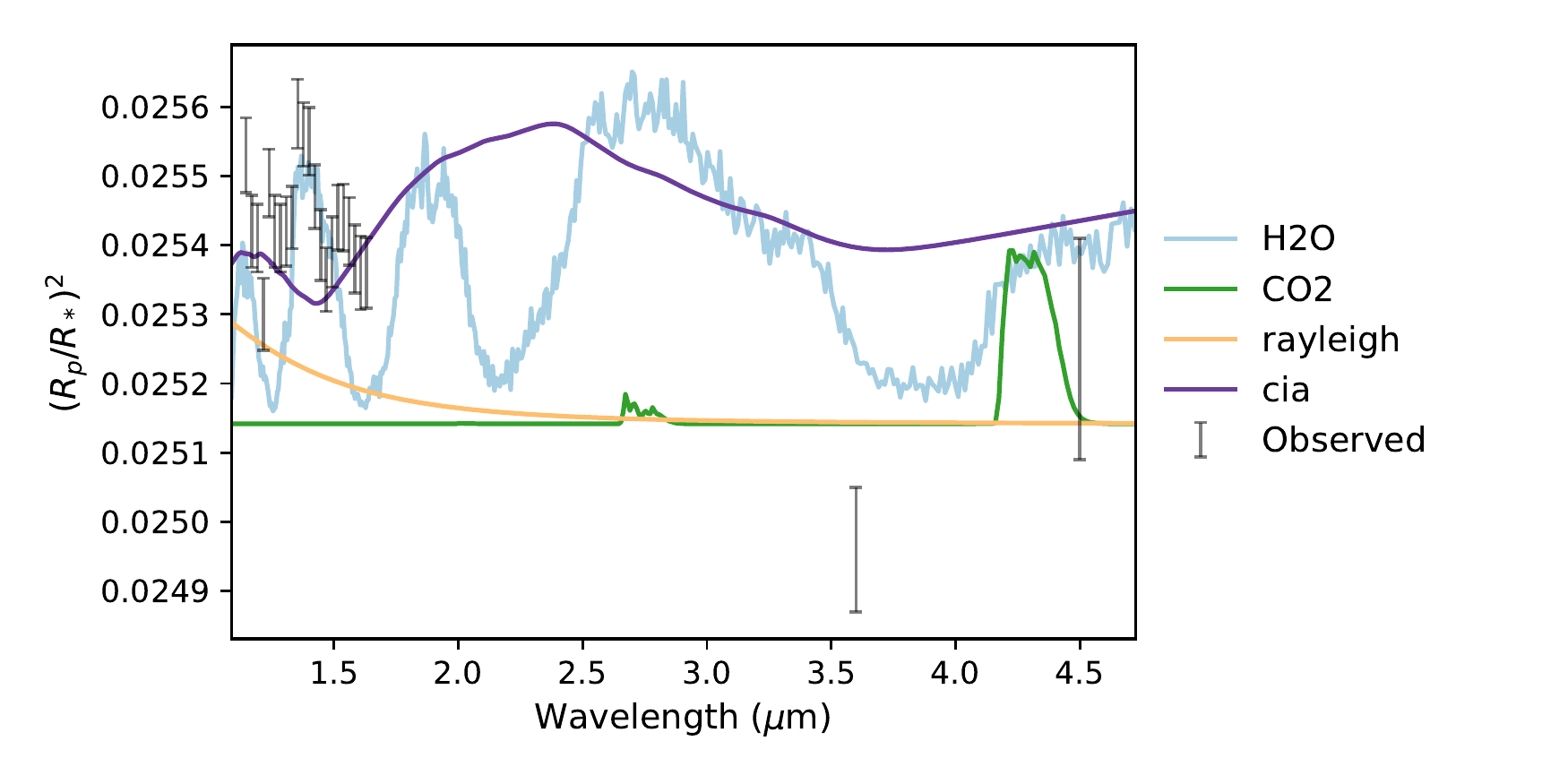}
        \caption{Tau-REx transmission retrieval results, including Spitzer points from \cite{14BlHaMa.wasp43b}, with H$_2$O and CO$_2$ (top) and the contributions of each molecule to the retrieval result (bottom). The different shading in the first panel corresponds to 1~$\sigma$ and 2~$\sigma$ regions. }
        \label{fig:TauRex_spitzer_CO2}
\end{figure}

We ran some retrievals including the Spitzer points of \cite{14BlHaMa.wasp43b}. Figure~\ref{fig:TauRex_spitzer} shows the resulting cloud-free retrieval and molecular contributions for AlO and H$_2$O only. Figure~\ref{fig:TauRex_spitzer_CO2} shows  the resulting cloud-free retrieval and molecular contributions for CO$_2$ and H$_2$O only. 
The model with H$_2$O and AlO is very strongly preferred over that with H$_2$O and CO$_2$; it gives a Bayes Factor of 12.4, corresponding to a significance of higher than 5~$\sigma$. A model with H$_2$O and AlO gives a similar Bayes factor of 12.0 when compared to one with H$_2$O only. We tried all the molecules mentioned above and did not find any evidence to include any of them. We arrive at the same conclusion when using the Spitzer points from \cite{19MoDaDi.wasp43b}, and when replacing CO$_2$ with CO \citep{15LiGoRo.CO}. 

\begin{figure}[]
        \centering
        \includegraphics[width=0.45\textwidth]{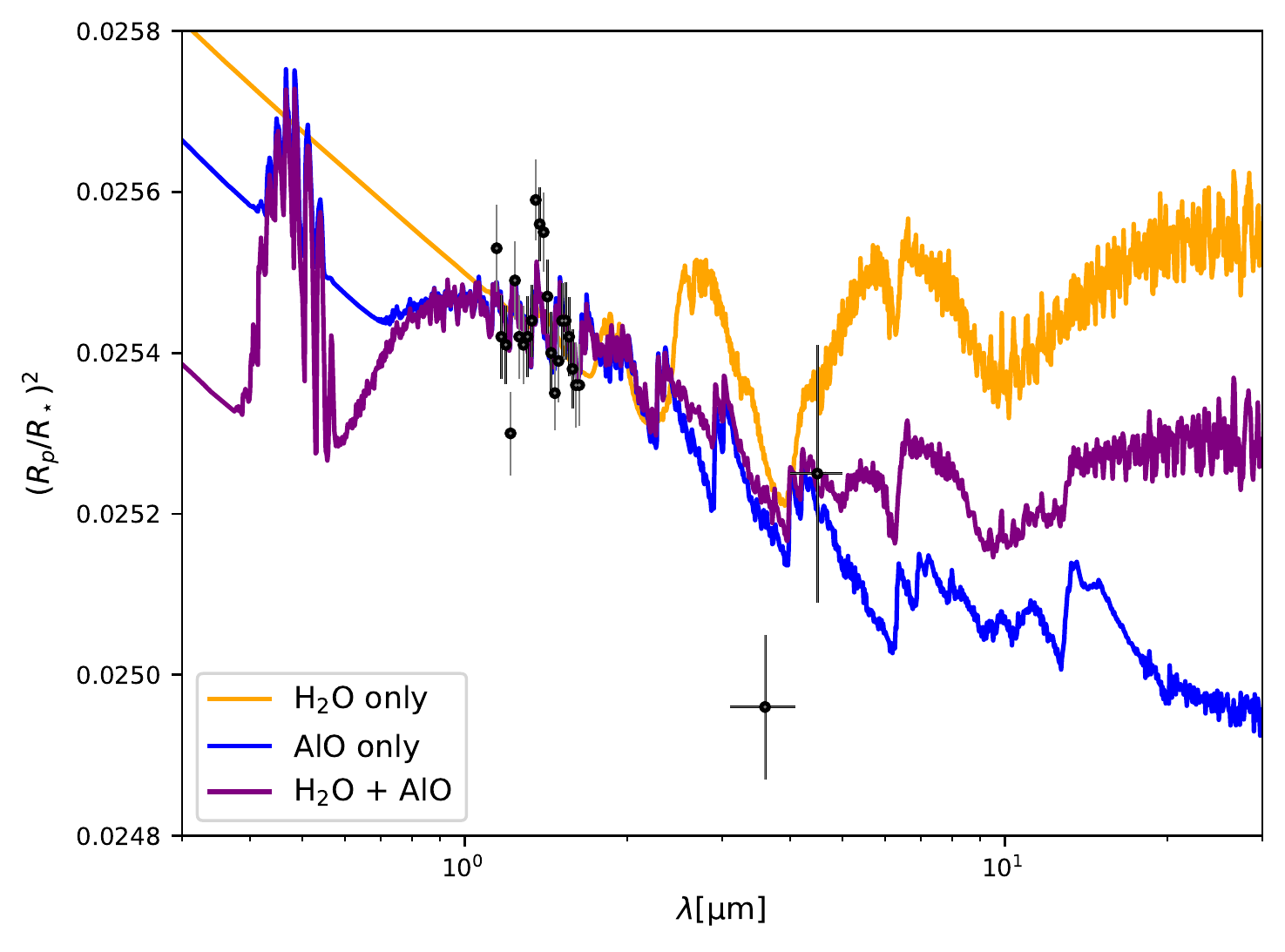}
        \includegraphics[width=0.45\textwidth]{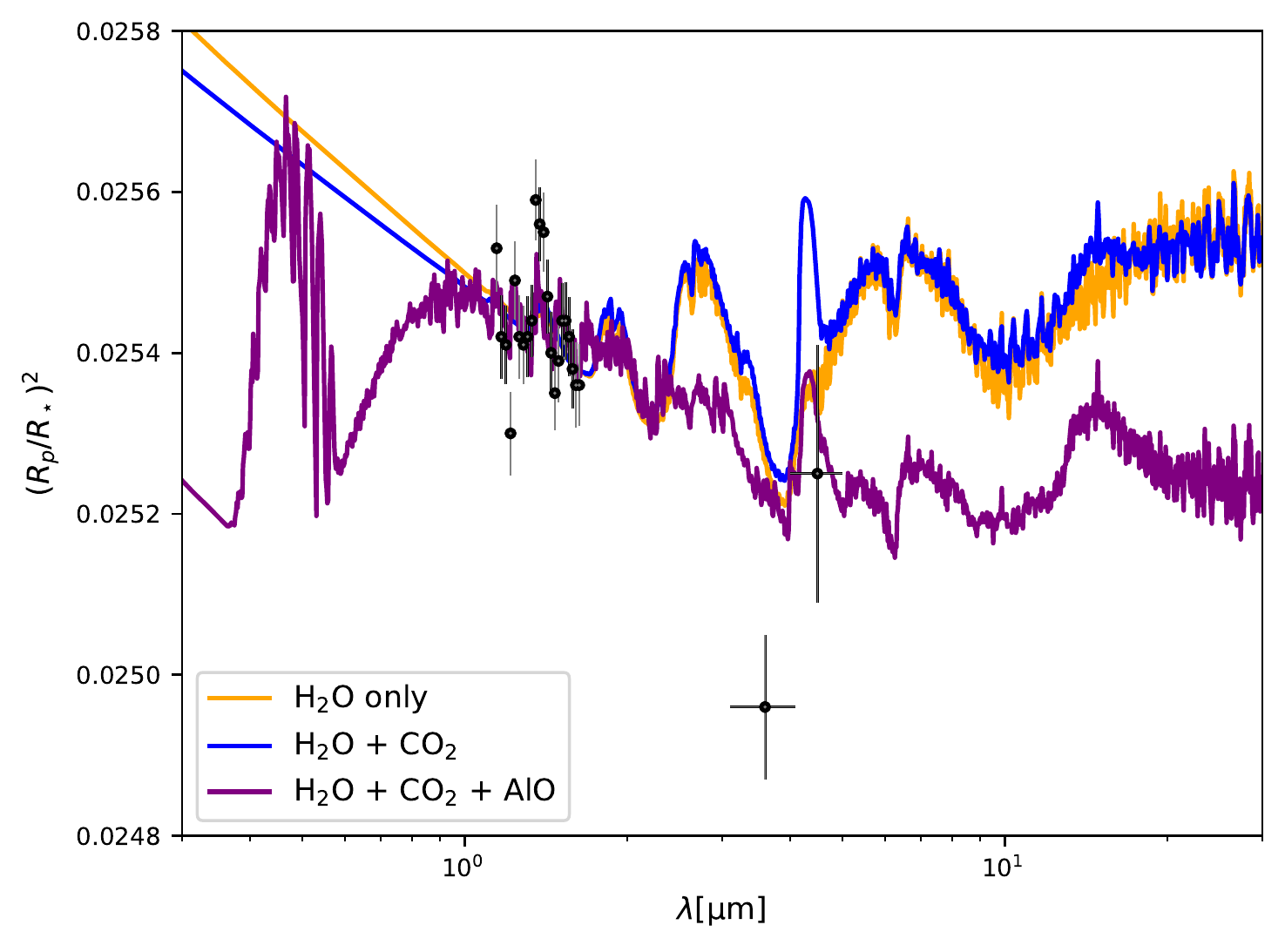}
        \caption{Best fit forward models between 0.3~-~30~$\mu$m, from our retrievals including both HST WFC3 and Spitzer IRAC  data, indicating some absorption features which could be identified from transmission spectra of future missions such as JWST or ARIEL. Top: H$_2$O  only, AlO only, and H$_2$O + AlO. Bottom: H$_2$O only, H$_2$O + CO$_2$, and H$_2$O + CO$_2$ + AlO.
                The HST WFC3 and Spitzer IRAC data are included for reference. 
         }
        \label{fig:ARCiS_JWST}
\end{figure}

Some ground-based observations of WASP-43b have also been made, with available data in the optical region from ground-based instruments 
from \cite{14MuPaZa.wasp43b} and \cite{19WeLoMe.wasp43b}, and broad-band data from \cite{14ChBoWa.wasp43b} and \cite{18VaGaVa.exo}; we also did not use them for our retrievals, due to their large uncertainties and issues with combining data from different instruments.
The most notable AlO absorption feature can be seen in the top panel of Figure~\ref{fig:ARCiS_JWST}, at around 0.4~-~0.5~$\mu$m. More accurate observations in this region would help confirm our findings and constrain AlO abundances. Evidence was found by \cite{14MuPaZa.wasp43b} for the Na I doublet around 589~nm at around 2.9~$\sigma$ confidence, but no evidence of K. \cite{19WeLoMe.wasp43b} recently found no evidence of either Na or K. The data of both studies do not cover the region of AlO absorption at around 0.4~-~0.5~$\mu$m.  

Figure~\ref{fig:ARCiS_JWST} illustrates some absorption features in the wavelength region of JWST (0.6~-~28~$\mu$m) and ARIEL (0.5~-~7.8~$\mu$m), which could help confirm our detections and constrain the molecular abundances and other retrieved parameters, which we note cannot be constrained by WFC3 data alone. For simplicity, we compare best fit forward models, based on retrievals using H$_2$O only, AlO only, H$_2$O + AlO, and H$_2$O only, H$_2$O + CO$_2$, H$_2$O + CO$_2$ + AlO. Unfortunately, one of the most prominent absorption features of AlO occurs just below 0.5~$\mu$m (see the top panel of Figure~\ref{fig:ARCiS_JWST}), which would therefore not be observable by either JWST or ARIEL. The Twinkle space telescope, however, which is due for launch in early 2022, has two spectrometers (visible, 0.4~-~1~$\mu$m, and infrared, 1.3~-~4.5~$\mu$m)~\citep{19EdRiZi.twinkle}, and so should be able to observe this feature. The Hubble STIS instrument is currently available, with an observational wavelength region which also covers that of the strong AlO absorption feature. 

\subsection{Opacity data requirements} \label{sec:opacity}

Of the above-mentioned Al-containing species that are expected from chemical equilibrium processes (see Section~\ref{sec:eqchem}), opacity line list data in the IR region covered by HST and spitzer data exist for AlH~\citep{18YuWiLe.AlH} and AlO~\citep{15PaYuTe.AlO} only. There are no significant absorption features in the 1.1~-~1.7~micron region for AlH, but there is some absorption in the region where the Spitzer data covers, around 3.6~microns. There is data available from MoLLIST~\citep{MOLLIST}, which can be found in ExoMol format~\citep{jt790} on the ExoMol website, for AlF and AlCl~\citep{18YoBe.exo}. This data, however, only extends up to around 2350~cm$^{-1}$ for AlCl and 3900~cm$^{-1}$ for AlF (i.e. it does not cover below 4.2 or 2.5~microns for AlCl and AlF, respectively). 
There is currently no line list data available in the literature for Al$_2$O or AlOH. 

Regarding the species mentioned in Section~\ref{sec:spitzer}, which could potentially be used to explain the Spitzer data points, the line list data for SiO~\citep{ExoMol_SiO}, AlF~\citep{18YoBe.exo}, and CaF~\citep{18HoBe.CaF} are currently not computed up to a high enough energy (relating to a low enough wavelength) to cover the region of the WFC3 data, or into the visible. The line list data for NS~\citep{18YuBoGo.NS}, PH~\citep{19LaTeYu.PH}, and PS~\citep{17LaPaLo.PO} do already cover the WFC3 wavelength region; however, there are no significant absorption features that can be used to detect these molecules in this region.  The ExoMol group~\citep{ExoMol_gen,jt804} are working on theoretical calculations that would extend some of these line lists, which will aid future investigations


This work shows that Al$_2$O or AlOH, along with AlCl and AlF, could be interesting molecules to have opacity data for in the IR region, in order to facilitate more in-depth studies into aluminium bearing atmospheres in the future. This would be particularly useful in the era of JWST~\citep{06GaMaCl.JWST} and ARIEL~\citep{18PaBeBa.ARIEL}. \cite{18TeYu.exo} give a good summary of the line lists available from ExoMol (as of 2018; the project is ongoing, with periodic additions of new molecular line lists and updates to existing ones), and of their level of completeness and data coverage. 

\subsection{Comparison to previous work}\label{sec:compare}

 Our simplest model includes H$_2$O and AlO only and is cloud-free. We can compare the retrieved parameters that are given in Figure~\ref{fig:TauRex_post1} to previous studies of the same transmission spectra. We retrieve a temperature of $858^{+419}_{-289}$~K, compared to $640^{+145}_{-129}$~K from \cite{14KrBeDe.wasp43b}. Another study of 30 exoplanets from \cite{18TsWaZi.exo} yielded $957\pm343$~K for WASP-43b, which used the transmission spectra generated from their own data reduction methods. This same data for WASP-43b was used by \cite{18FiHe.exo} in a recent study of 38 exoplanets; they retrieved a temperature of $835^{+340}_{-121}$~K. 

Our retrieved molecular abundances (within 1~$\sigma$ ranges) are 2.9$\times10^{-7}$~-~3.8$\times10^{-3}$ and 7.6$\times10^{-8}$~-~4.2$\times10^{-4}$ for H$_2$O and AlO, respectively. The water vapour volume mixing ratio was found to be 3.2$\times10^{-5}$~-~1.6$\times10^{-3}$ by \cite{14KrBeDe.wasp43b}, 1.1$\times10^{-6}$~-~1.7$\times10^{-2}$ by \cite{18FiHe.exo}, 3.5$\times10^{-7}$~-~5.5$\times10^{-3}$ by \cite{18TsWaZi.exo}, 3.6$\times10^{-5}$~-~3.9$\times10^{-2}$ by \cite{19WeLoMe.wasp43b}, and 1$\times10^{-4}$~-~1$\times10^{-3}$ by \cite{19IrPaTa.wasp43b}. We can only compare the retrieved H$_2$O abundance to previous studies, as AlO was not included in any previous retrievals.

While all the values are roughly in agreement, within the error bars, there are huge uncertainties on the retrieved parameters. We note that there is degeneracy between the retrieved radius, temperature, and molecular abundances. For example, a lower temperature can be compensated by a higher radius and higher abundances without much variation in the final spectra. This is sometimes referred to as the `normalisation degeneracy' \citep[see e.g.][]{12BeSe.exo,14Griffith.exo,17HeKi.exo,18FiHe.exo}. As our retrieval is based on HST WFC3 data alone, we are not able to place any tight constraints on the retrieved parameters. 
Figure~\ref{fig:TauRex_post1}  does, however, show that there is a positive correlation between H$_2$O and AlO, which means that the ratio of the two should be better constrained than the absolute abundances of each. We can use some approximations to compare this to solar abundances. Based on the solar elemental abundances given in ~\cite{09AsGrSa.sun}, solar log(Al/O)~=~-2.24. We can compare this to our retrieved value of log(AlO/(AlO+H$_2$O))~=~$-1.48^{+1.48}_{-1.67}$. This is assuming that most of the oxygen is in H$_2$O, and all the aluminium is contained within AlO. In reality, oxygen would likely also be locked into silicate and other species, although this would not be expected to have a significant effect. If Al were contained within other species, this would be lower than the actual abundance of Al/O, but if oxygen were contained in other species our estimate would be higher than the true value.  


\section{Summary}\label{sec:summary}

In this work we have performed a re-analysis of the HST WFC3 transmission spectrum of WASP-43b between 1.1~-~1.7$\mu$m. We have tested the statistical significance of including every molecule for which high-temperature opacity data exists in this wavelength region. We find strong evidence ($>$~5~$\sigma$) to justify the inclusion of AlO and H$_2$O in our model, but not for any other molecules in this region. We investigate the effects of including Spitzer IRAC data points at 3.6~$\mu$m and 4.5~$\mu$m, and find the inclusion of these points gives strength to our AlO and H$_2$O detection.
The presence of AlO at the temperatures and pressures that these transmission observations are probing is not expected from equilibrium chemistry; its presence is therefore evidence of disequilibrium processes in the atmosphere of WASP-43b. Detecting heavy elements such as Al in a planetary atmosphere also gives some insight into planet formation processes~\citep{12JoLi.exo,14HaHi.exo}. None of the previous studies analysing the transmission spectra of WASP-43b considered AlO as a potential molecule to include in their retrieval process, and only a small set of the molecules for which there is data available were included. It should be noted that the AlO line list was not available when the atmosphere of WASP-43b was first analysed by \cite{14KrBeDe.wasp43b}. Our approach of considering all molecules for which data currently exists highlights the importance of projects such as ExoMol and HITEMP for expanding our knowledge of high-temperature exoplanet atmospheres, in particular in the era of space missions such as JWST and ARIEL. We note that molecular abundances and other retrieved parameters cannot be accurately constrained by WFC3 data alone, as has previously been pointed out by works such as \cite{17HeKi.exo}, and observations across a wide range of wavelengths are essential for expanding on this and other works on WASP-43b.

\section*{Acknowledgements}
This project has received funding from the European Union's Horizon 2020 Research and Innovation Programme, under Grant Agreement 776403, and from the European Research Council (ERC) under the European Union's Horizon 2020 research and innovation programme under grant agreement No 758892, ExoAI. 
We thank the referee for their constructive comments on the manuscript. 

\bibliography{WASP_paper}
\bibliographystyle{aa}



\onecolumn
\begin{appendix}
        \section{}

\begin{table*}[h]
        \caption{Nested Sampling Global Log-Evidence (log($E$))
                 of various retrievals of the HST/WFC3 data, along with the natural log of the Bayes factor, ln$(B_{01})$, and $\sigma$ likelihood against: a flat `base' retrieval (i.e. with no molecular features), a retrieval with only H$_2$O included, and a retrieval with only AlO included. For reference, the Nested Sampling Global Log-Evidence of the base flat retrieval is 163.7, for the H$_2$O-only retrieval it is 173.2, and for the AlO-only retrieval it is 175.4. The values for $\sigma$ have been interpolated from Table~2 of \cite{08Trotta.stats}. }
        \label{t:bayes_factors_appendix} 
        \centering  
        \begin{tabular}{l | l | l | l | l}
                \hline\hline
                \rule{0pt}{3ex}Molecules included       &       log($E$)        & Clouds? & ln$(B_{01})$ &        $\sigma$ \\
                \hline
                \multicolumn{5}{l}{\rule{0pt}{3ex}Compared to flat model} \\
                \hline
                \rule{0pt}{3ex}H$_2$O &       173.2   & No  & 9.5 & 4.7         \\      
                H$_2$O &       172.9   & Yes & 9.2 & 4.6        \\      
                AlO    &       175.4  & No &  12.3 & $>$~5         \\      
                AlO    &       175.3   & Yes & 11.6 & $>$~5         \\      
                AlO + H$_2$O   &       177.4   & No &  13.7 & $>$~5              \\
                AlO + H$_2$O   &       176.7   & Yes & 13 & $>$~5            \\
                \hline
                \multicolumn{5}{l}{\rule{0pt}{3ex}Compared to H$_2$O-only model} \\
                \hline
                \rule{0pt}{3ex}AlO + H$_2$O    &       177.4   & No  & 4.2 & 3.4          \\      
                AlO + H$_2$O    &       176.7   & Yes & 3.8 & 3.2            \\      
FeH + H$_2$O    &       174.1   &       No      &       0.9     &       -       \\
TiH + H$_2$O    &       173.5   &       No      &       0.3     &       -       \\
CH + H$_2$O     &       173.4   &       No      &       0.2     &       -       \\
TiO + H$_2$O    &       173.3   &       No      &       0.1     &       -       \\
HeH$^+$ + H$_2$O        &       173.3   &       No      &       0.1     &         -       \\
H$_2$CO + H$_2$O        &       173.2   &       No      &       0       &         -       \\
ScH + H$_2$O    &       173.2   &       No      &       0       &       -       \\
K + H$_2$O      &       173.2   &       No      &       0       &       -       \\
Na + H$_2$O     &       173.2   &       No      &       0       &       -       \\
CO$_2$ + H$_2$O &       173.0   &       No      &       -0.2    &       -       \\
NH$_3$ + H$_2$O &       172.8   &       No      &       -0.4    &       -       \\
C$_2$H$_2$ + H$_2$O     &       172.8   &       No      &       -0.4    &         -       \\
HCN + H$_2$O    &       172.8   &       No      &       -0.4    &       -       \\
OH + H$_2$O     &       172.8   &       No      &       -0.4    &       -       \\
MgO + H$_2$O    &       172.8   &       No      &       -0.4    &       -       \\
H$_2$S + H$_2$O &       172.7   &       No      &       -0.6    &       -       \\
HNO$_3$ + H$_2$O        &       172.7   &       No      &       -0.6    &         -       \\
CH$_4$ + H$_2$O &       172.6   &       No      &       -0.6    &       -       \\
C$_2$H$_4$ + H$_2$O     &       172.5   &       No      &       -0.7    &         -       \\
VO + H$_2$O     &       172.1   &       No      &       -1.1    &       -       \\
                \hline
                \multicolumn{5}{l}{\rule{0pt}{3ex}Compared to AlO-only model} \\
                \hline
\rule{0pt}{3ex}AlO + H$_2$O    &       177.4   & No  & 2.0 & 2.6       \\      
AlO + H$_2$O    &       176.7   & Yes & 1.3 & 2.2            \\
                \hline \hline
        \end{tabular}
\end{table*}

\begin{table*}[h]
        \caption{Nested Sampling Global Log-Evidence (log($E$))
                of various retrievals of the HST/WFC3 data, with various combinations of including Rayleigh scattering, collision induced absorption (CIA) of H$_2$-He and H$_2$-H$_2$ \citep{HITRAN_2016,01BoJo.cia}, or clouds. Results are shown for retrievals with only H$_2$O included, and retrievals with H$_2$O + AlO included. }
        \label{t:bayes_factors_cia_R} 
        \centering  
        \begin{tabular}{l | l | l | l }
                \hline\hline
                \rule{0pt}{3ex}CIA? & Rayleigh scattering? & Clouds? & log($E$)  \\
                \hline
                \multicolumn{4}{l}{\rule{0pt}{3ex}H$_2$O-only model}  \\
                \hline
                \rule{0pt}{3ex}No & No & No & 174.5\\
         No & Yes &  No & 175.2 \\
         Yes & No &  No & 173.2 \\
         Yes &  Yes &   No  & 173.2      \\
         No & No & Yes & 173.3\\
         No & Yes &  Yes & 173.7 \\
                 Yes & No &  Yes & 172.9  \\
                 Yes &  Yes &   Yes      & 172.9 \\
                 \hline
                 \multicolumn{4}{l}{\rule{0pt}{3ex}H$_2$O + AlO model} \\
                 \hline
                 \rule{0pt}{3ex}No & No & No & 177.6 \\
                 No & Yes &  No & 177.8 \\
                 Yes & No &  No & 177.4 \\
                 Yes &  Yes &   No  &   177.3 \\
                 No & No & Yes & 176.5    \\
                 No & Yes &  Yes & 176.5 \\
                 Yes & No &  Yes & 176.6 \\
                 Yes &  Yes &   Yes      & 176.7 \\
                \hline \hline
        \end{tabular}
\end{table*}

\begin{figure}[]
        \includegraphics[width=0.5\textwidth]{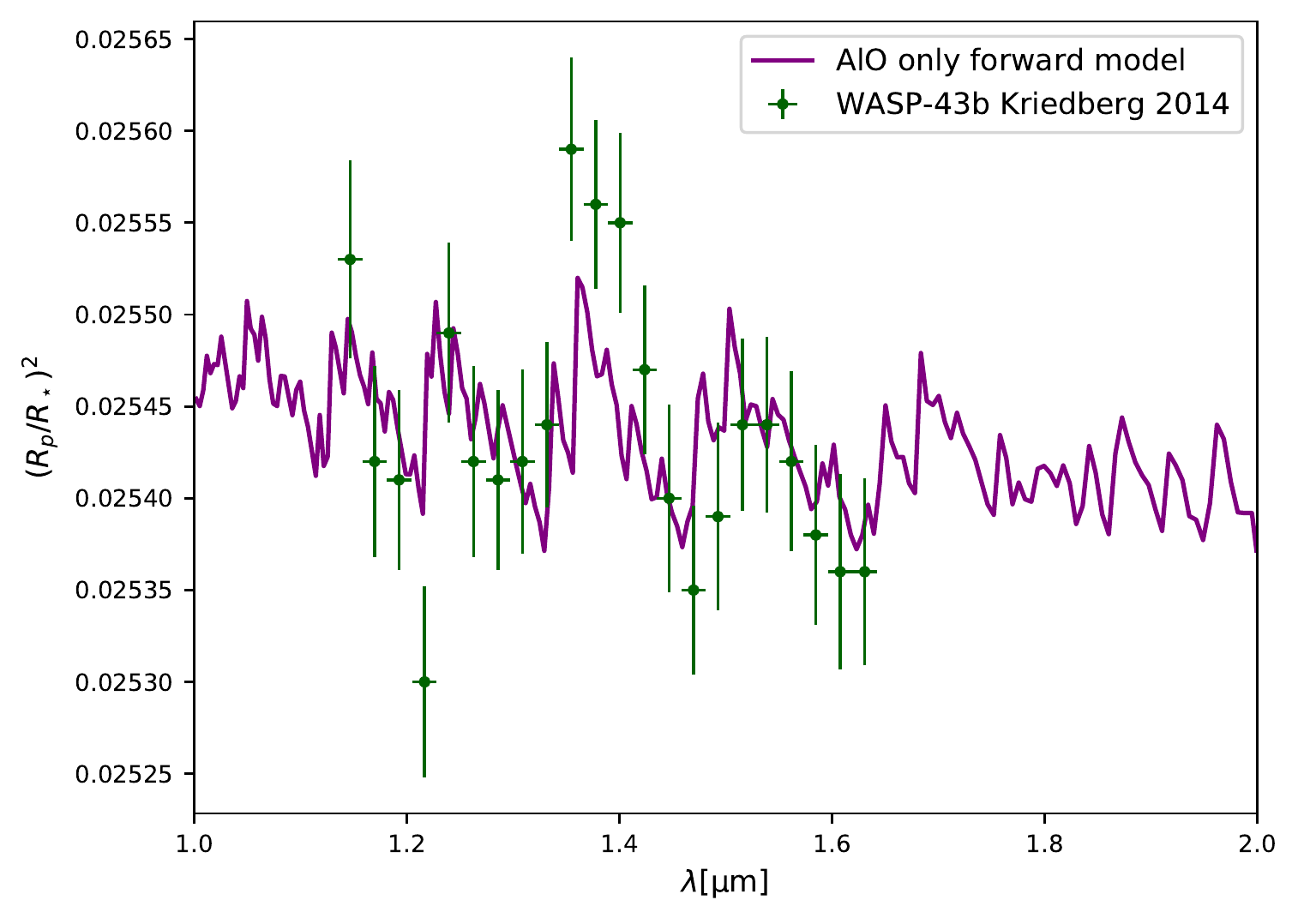}
        \includegraphics[width=0.5\textwidth]{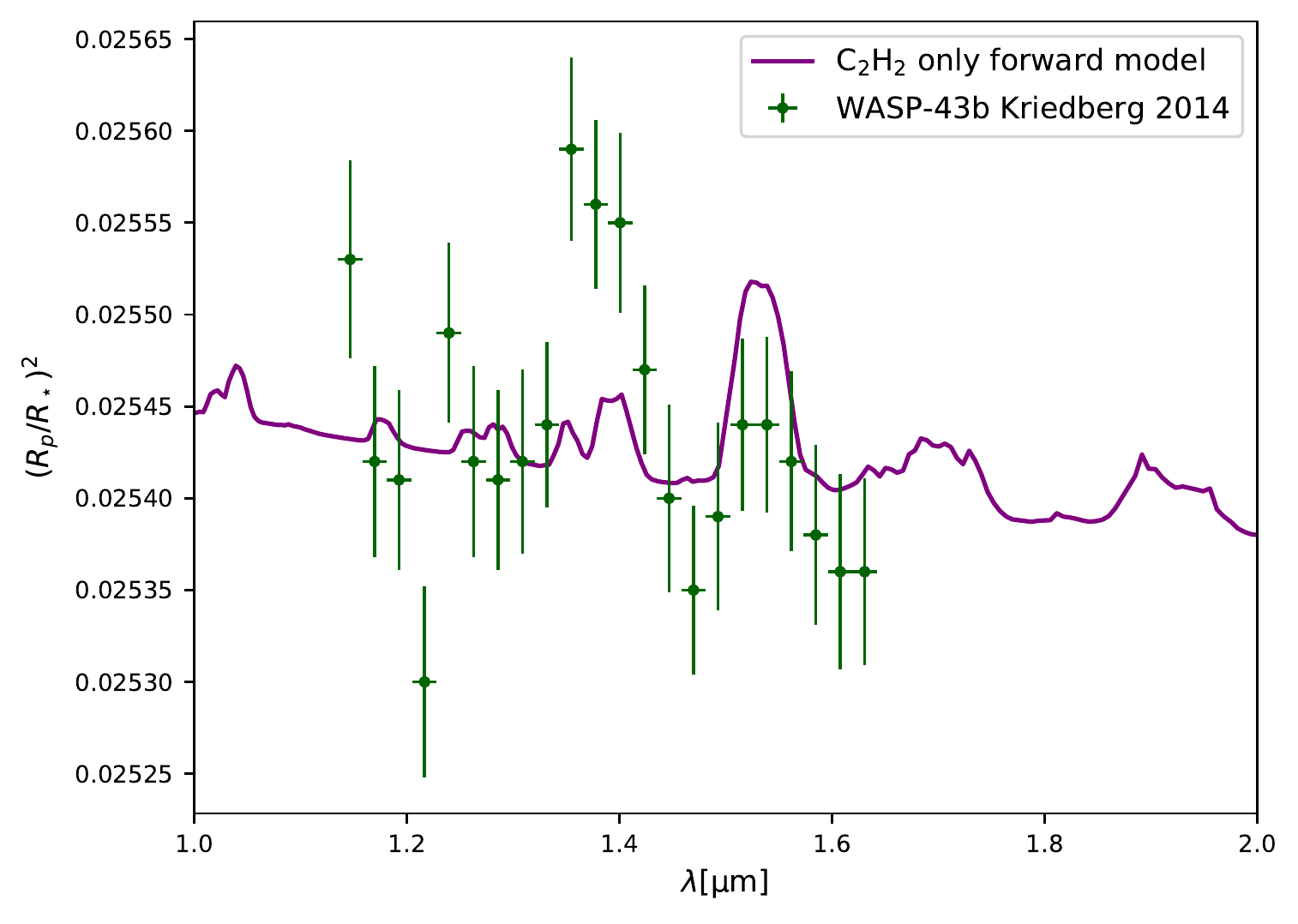}
        \includegraphics[width=0.5\textwidth]{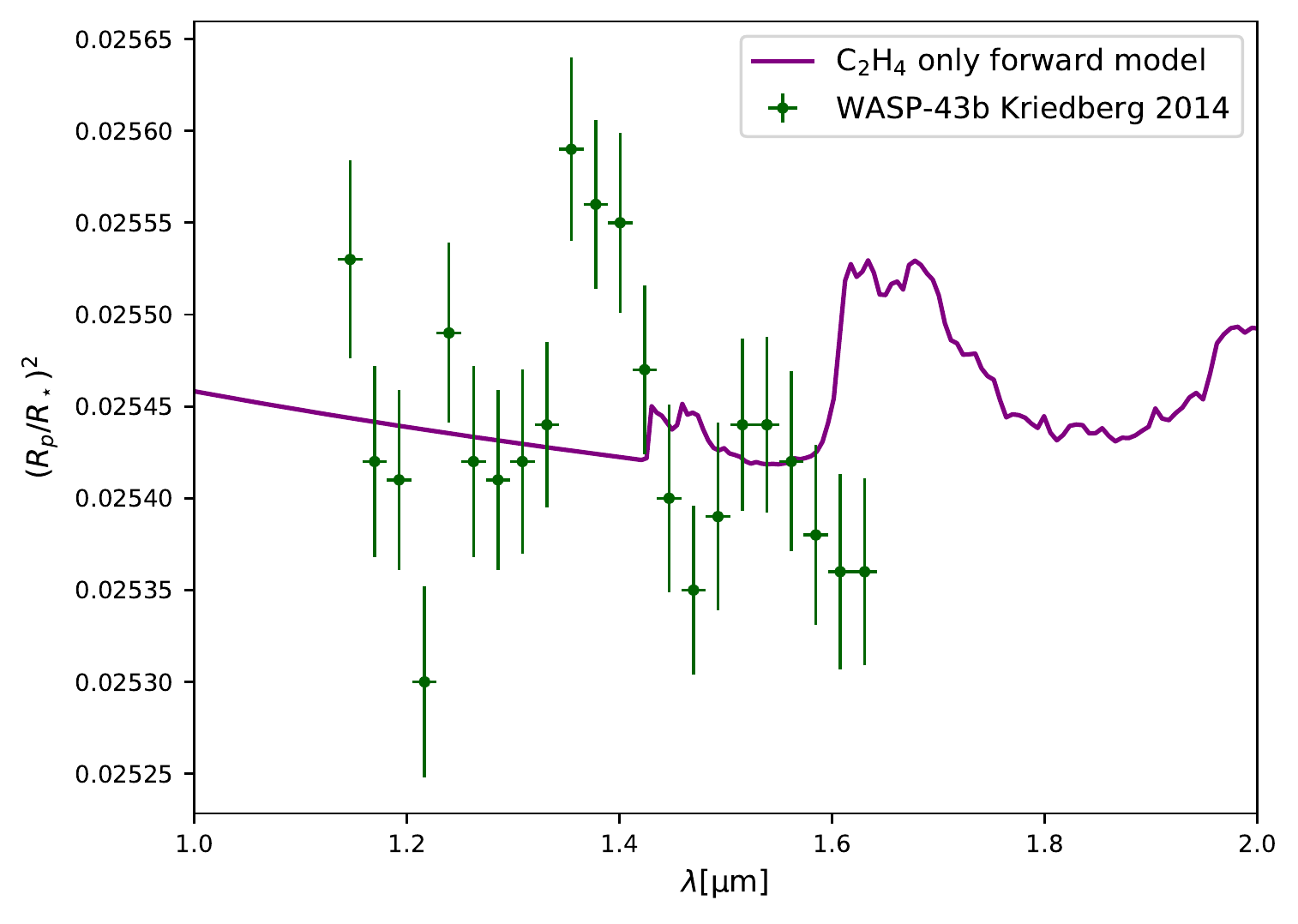}
        \includegraphics[width=0.5\textwidth]{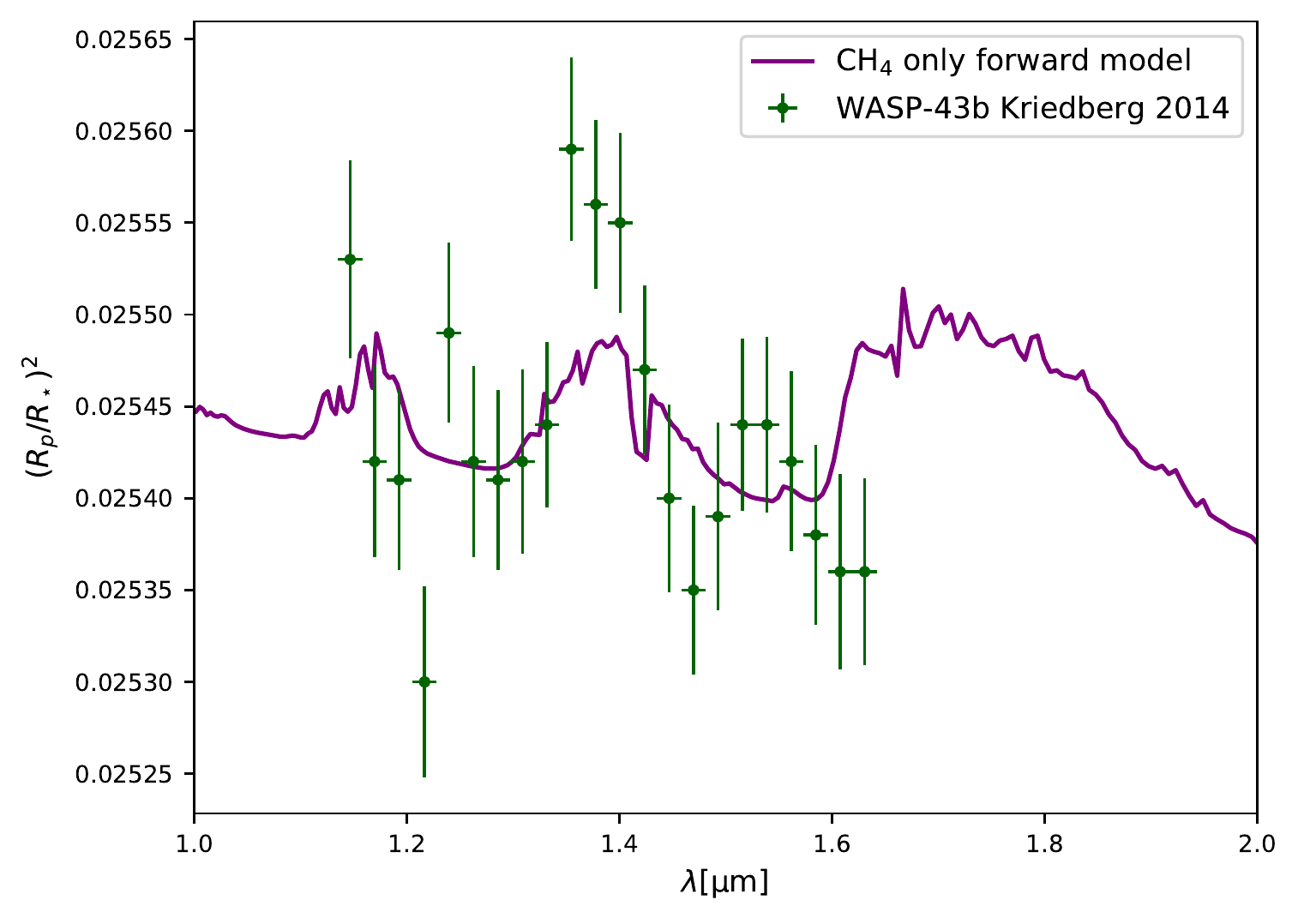}
        \includegraphics[width=0.5\textwidth]{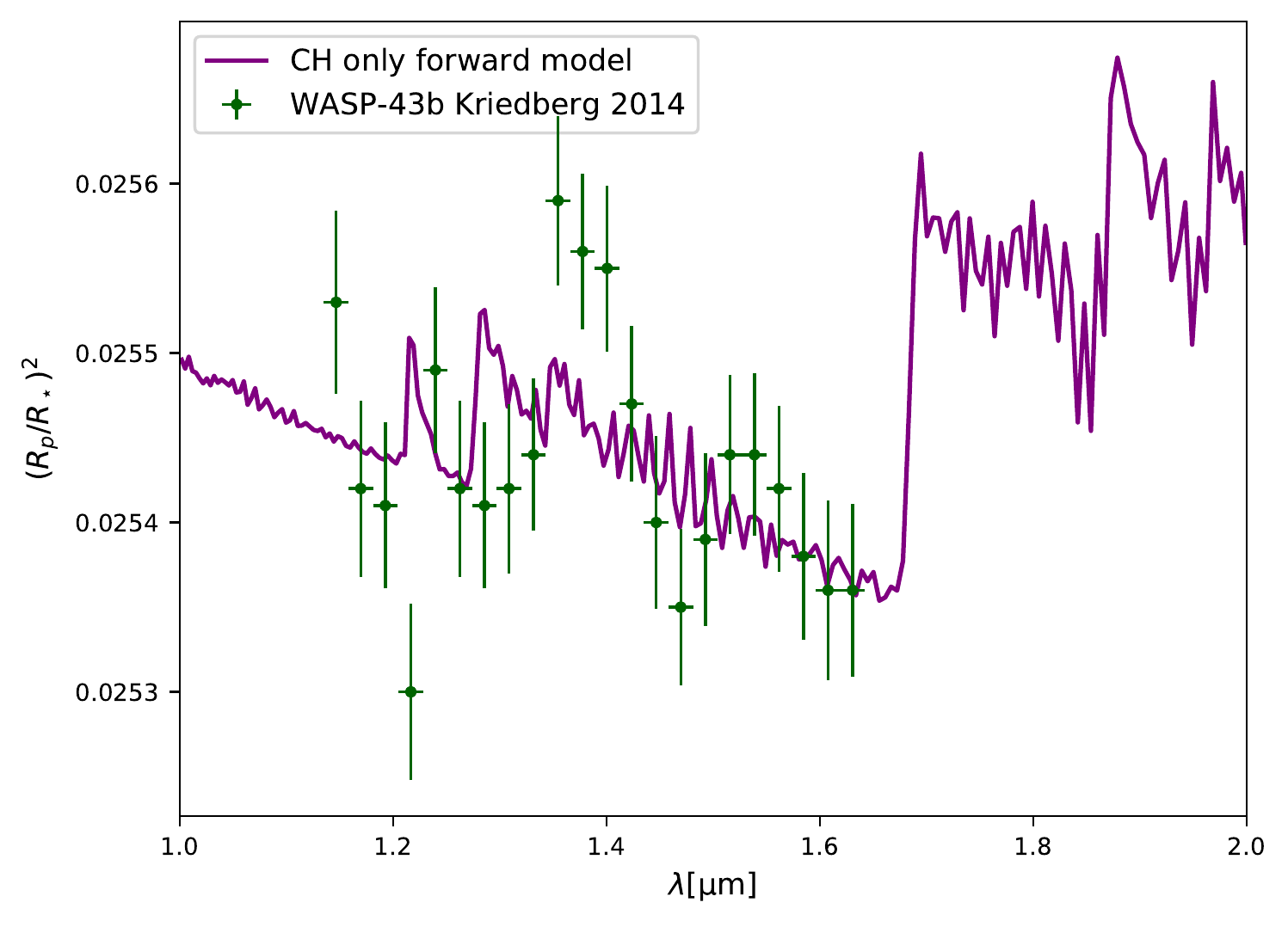}
        \includegraphics[width=0.5\textwidth]{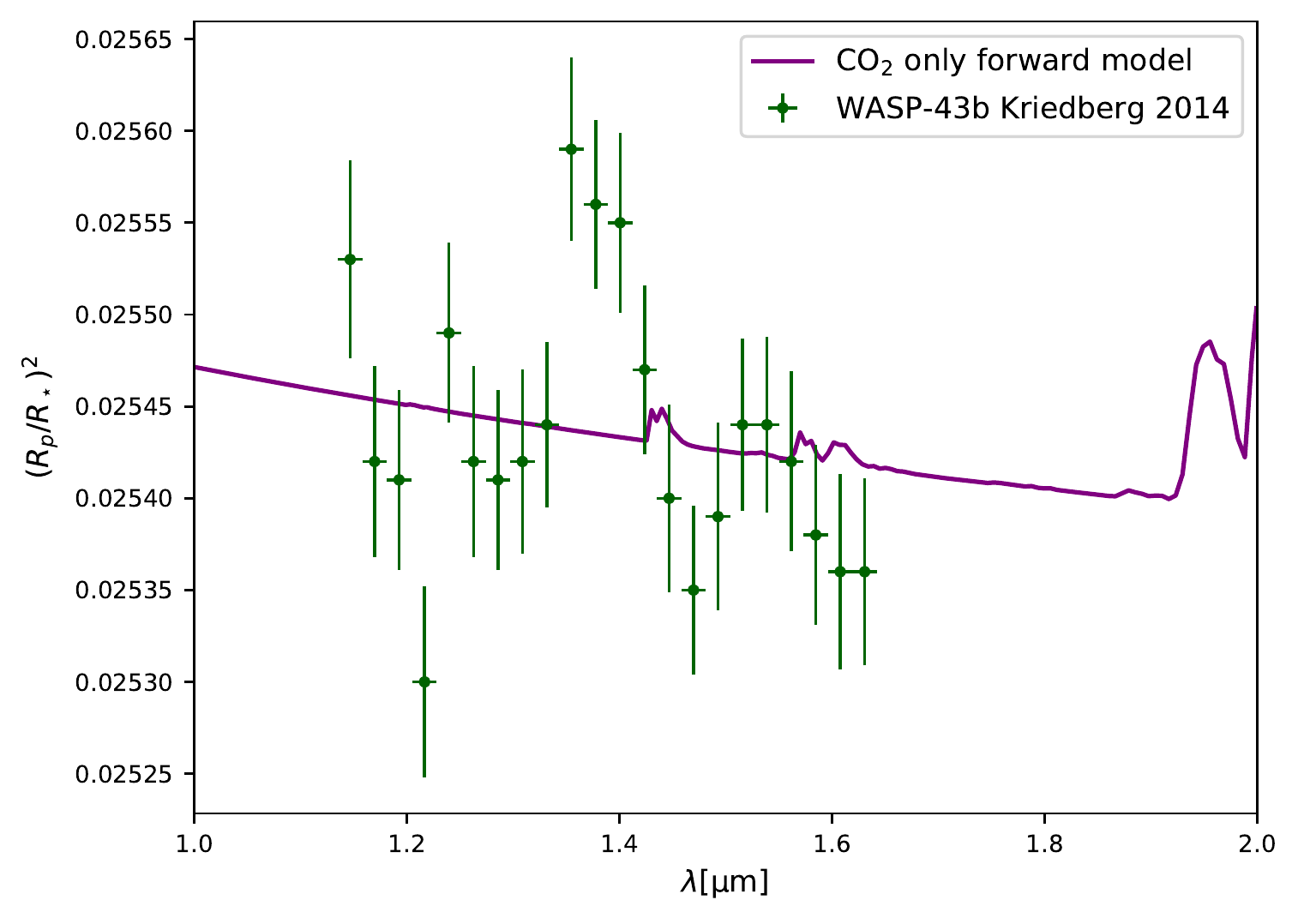}
        \caption{ARCiS forward models, each including one individual species, plotted alongside the transmission data for WASP-43b from \cite{14KrBeDe.wasp43b} in order to help assess which molecules to include in subsequent retrievals.}
        \label{fig:ARCiS_assess}
\end{figure}

\begin{figure}\ContinuedFloat
        \includegraphics[width=0.5\textwidth]{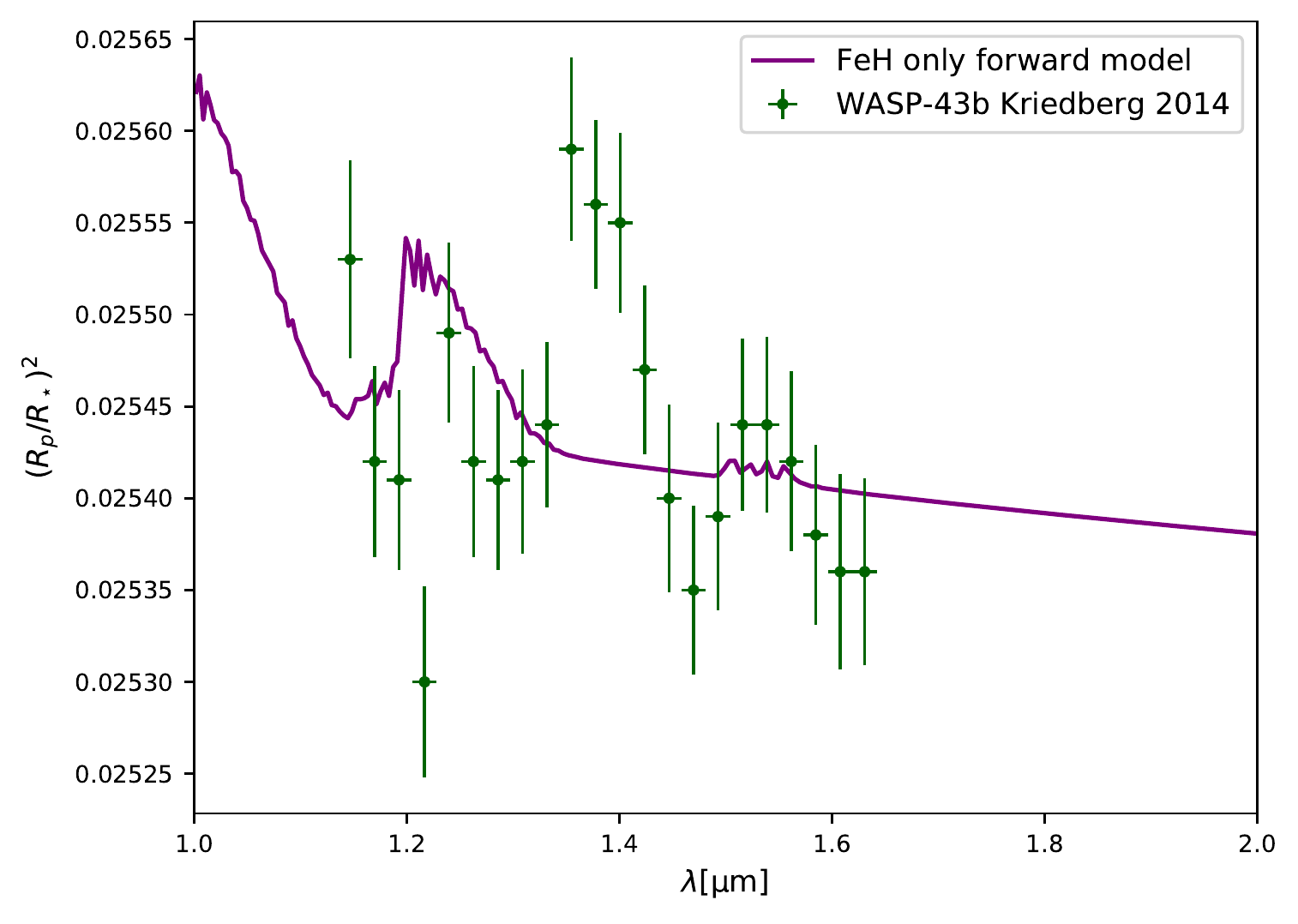}
        \includegraphics[width=0.5\textwidth]{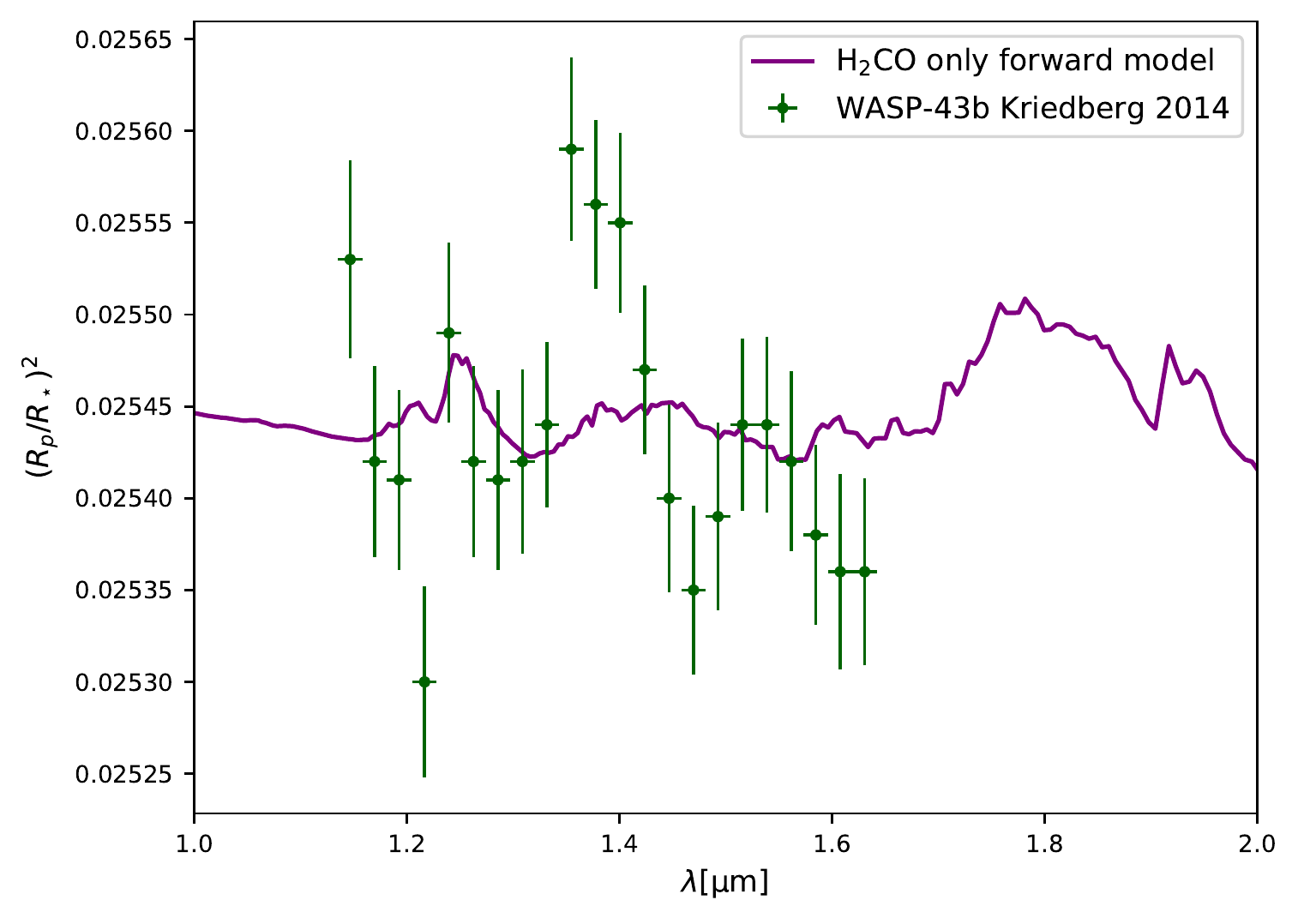}
        \includegraphics[width=0.5\textwidth]{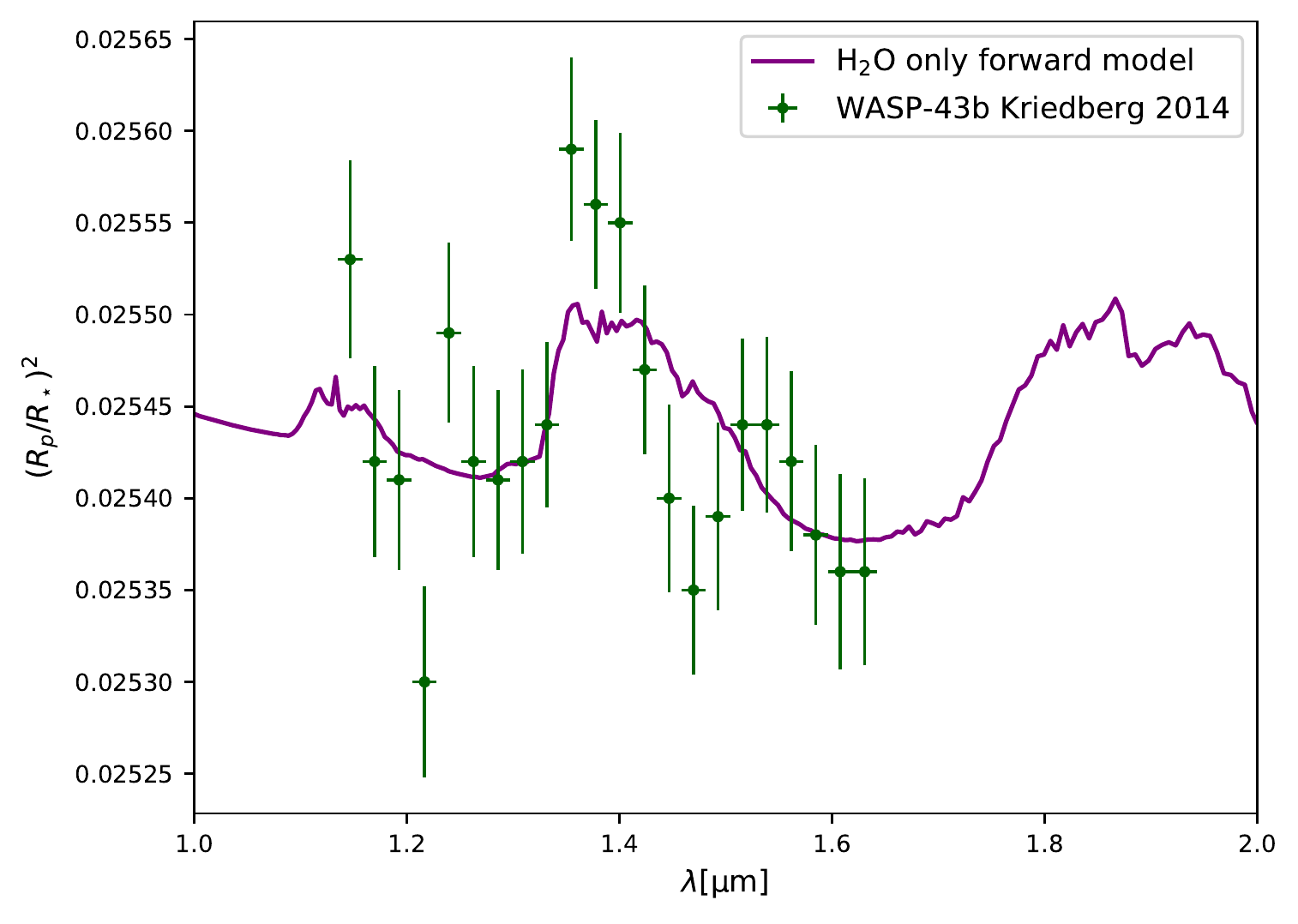}
        \includegraphics[width=0.5\textwidth]{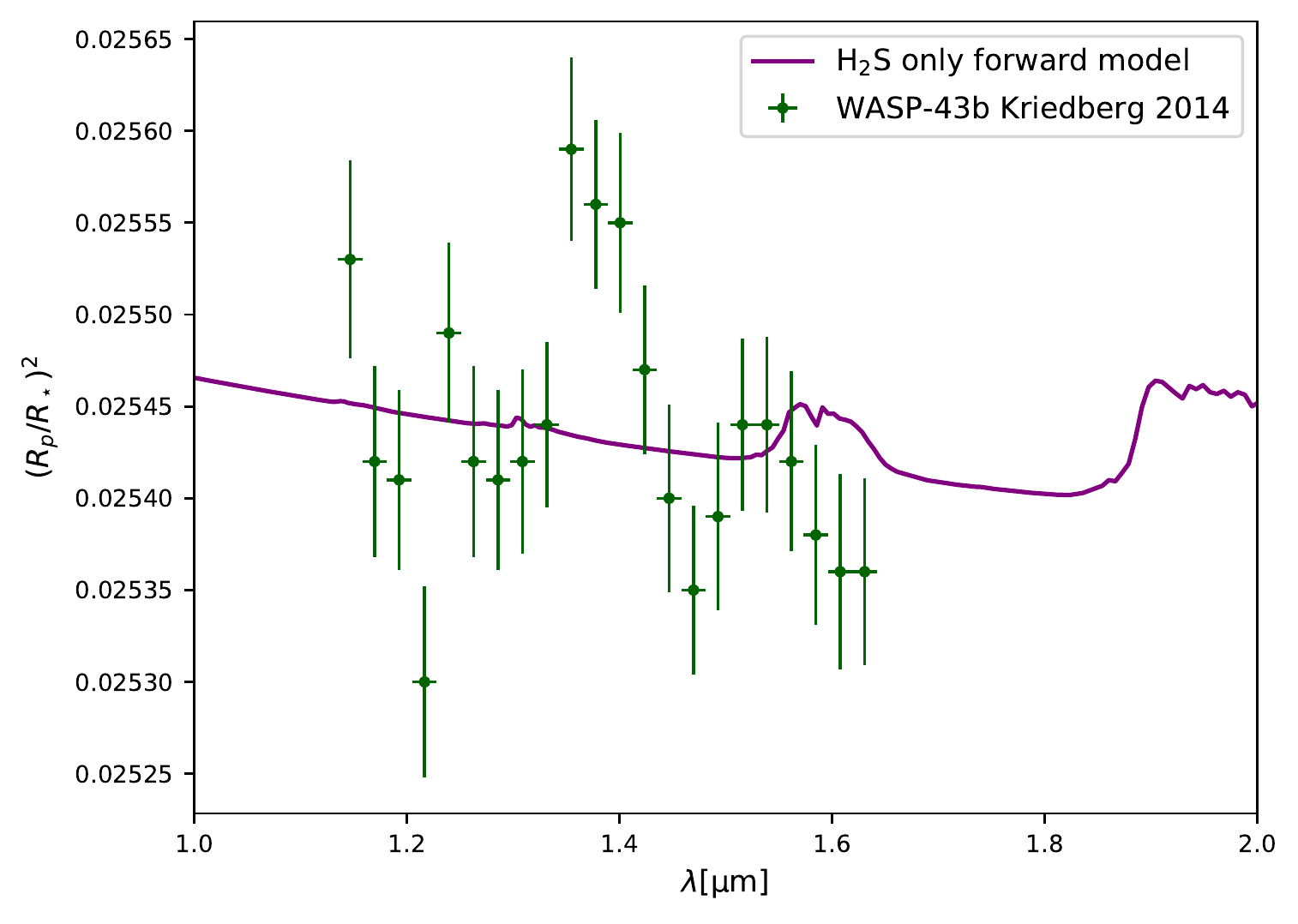}
\includegraphics[width=0.5\textwidth]{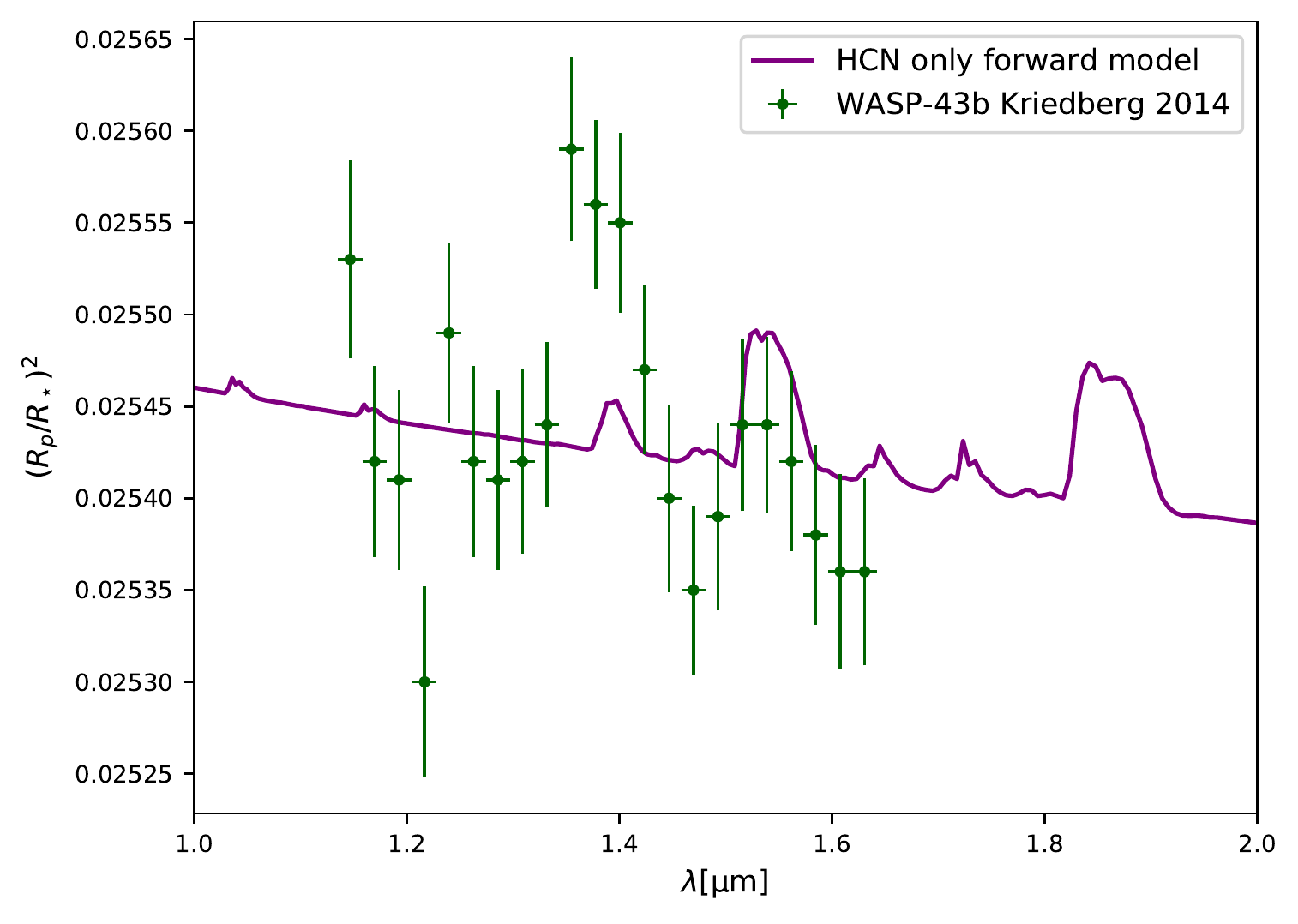}
\includegraphics[width=0.5\textwidth]{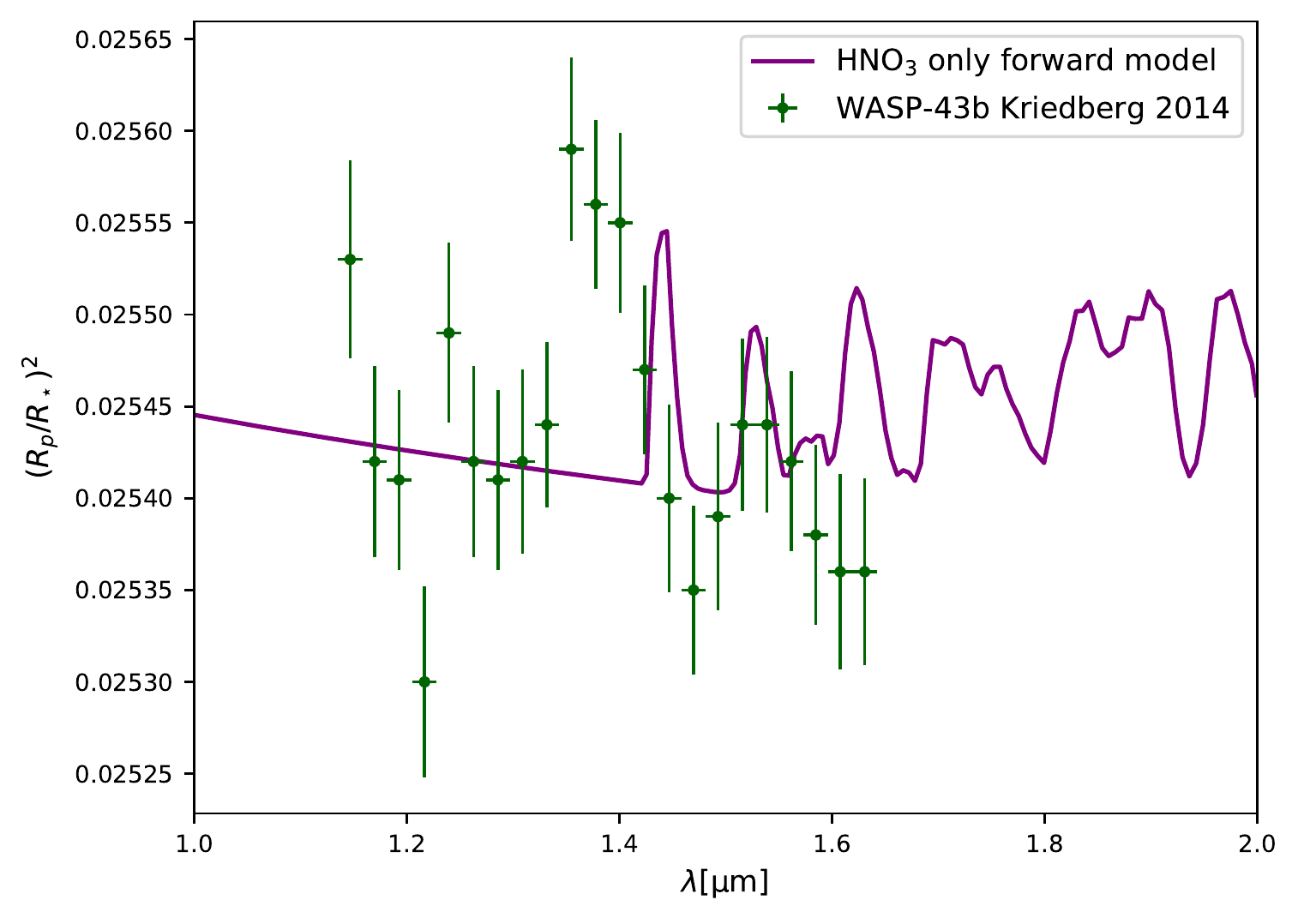}
        \caption{(continued) ARCiS forward models, each including one individual species, plotted alongside the transmission data for WASP-43b from \cite{14KrBeDe.wasp43b} in order to help assess which molecules to include in subsequent retrievals.} 
        \label{fig:ARCiS_assess}
\end{figure}

\begin{figure}\ContinuedFloat
        \includegraphics[width=0.5\textwidth]{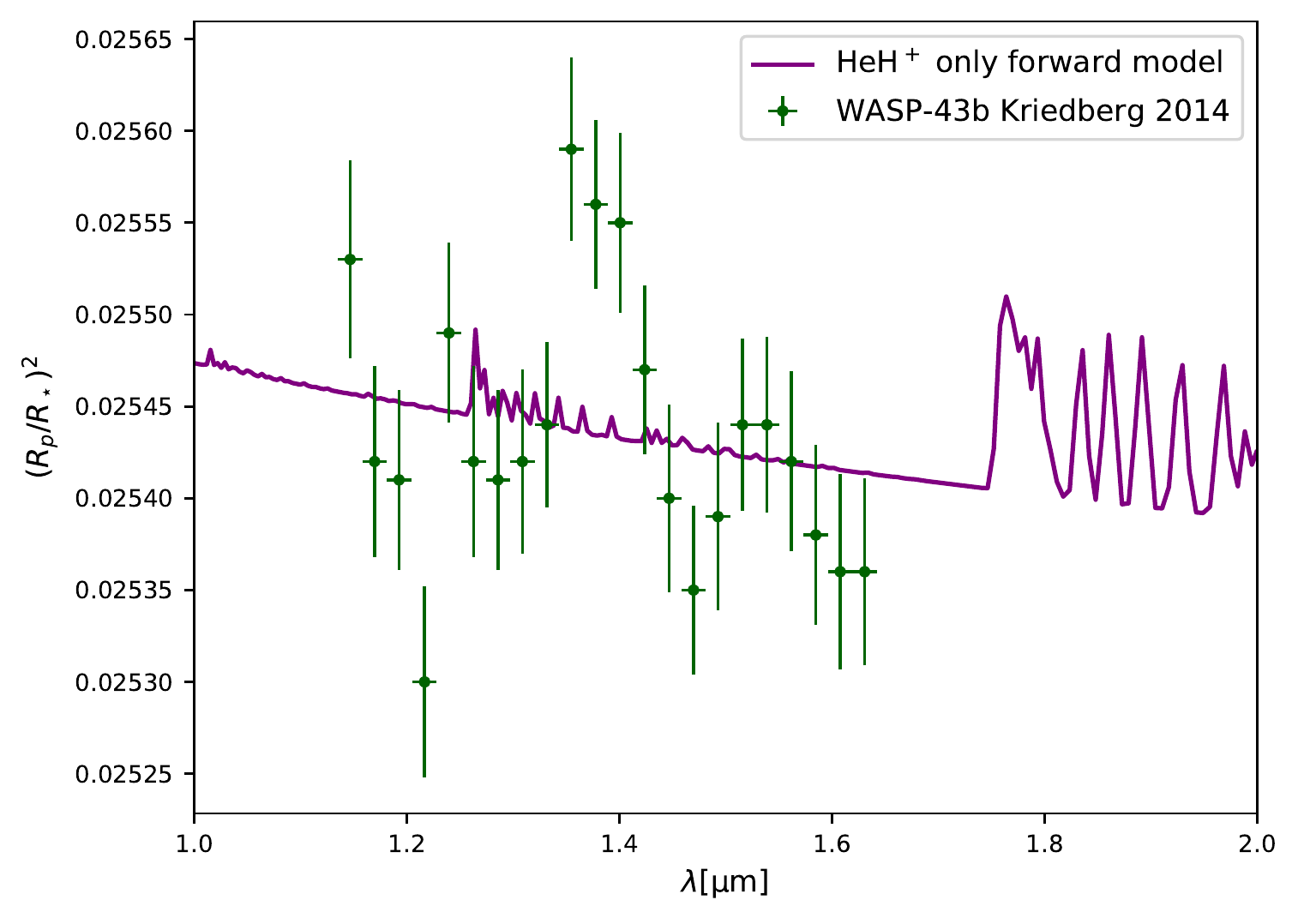}
        \includegraphics[width=0.5\textwidth]{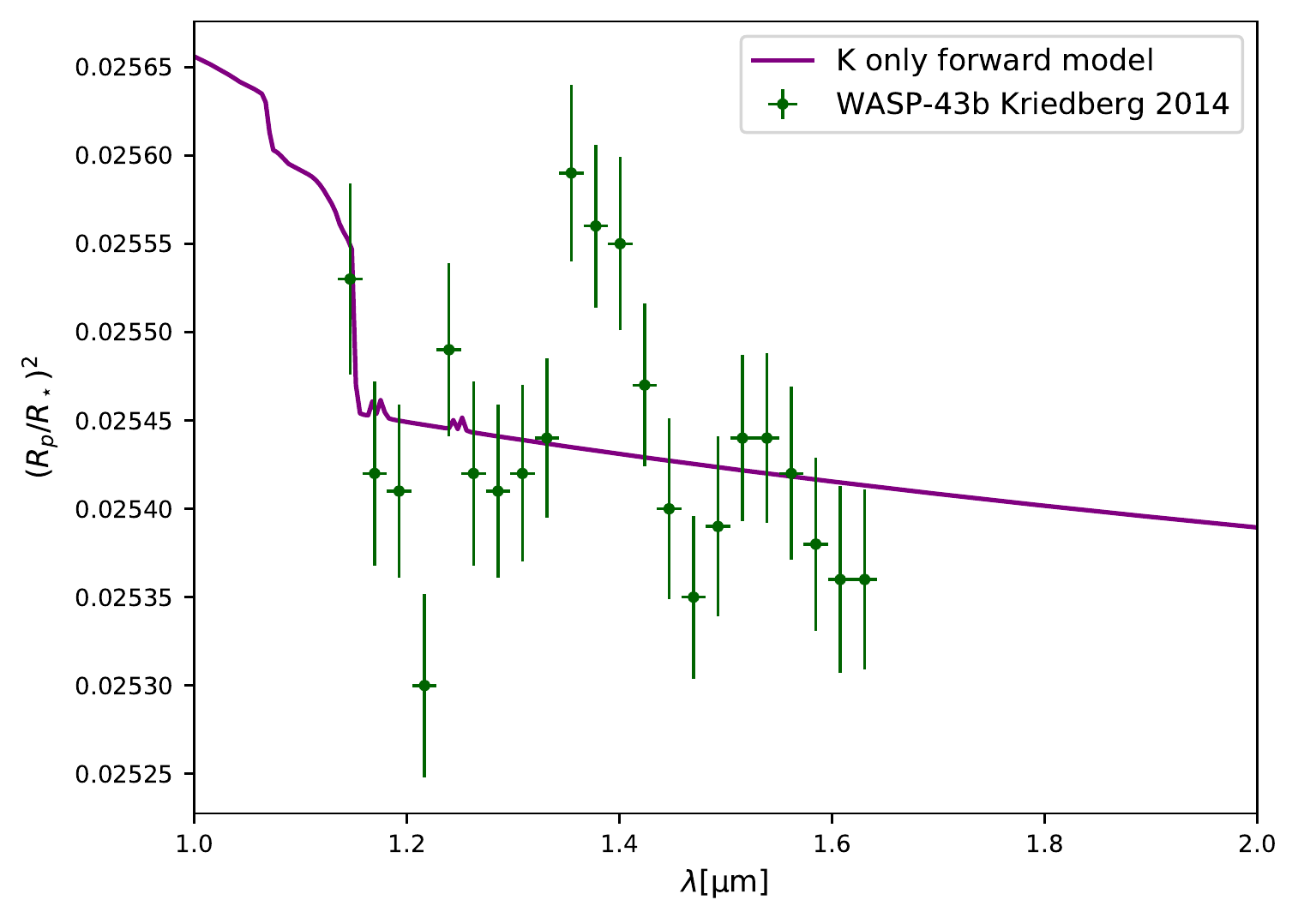}
        \includegraphics[width=0.5\textwidth]{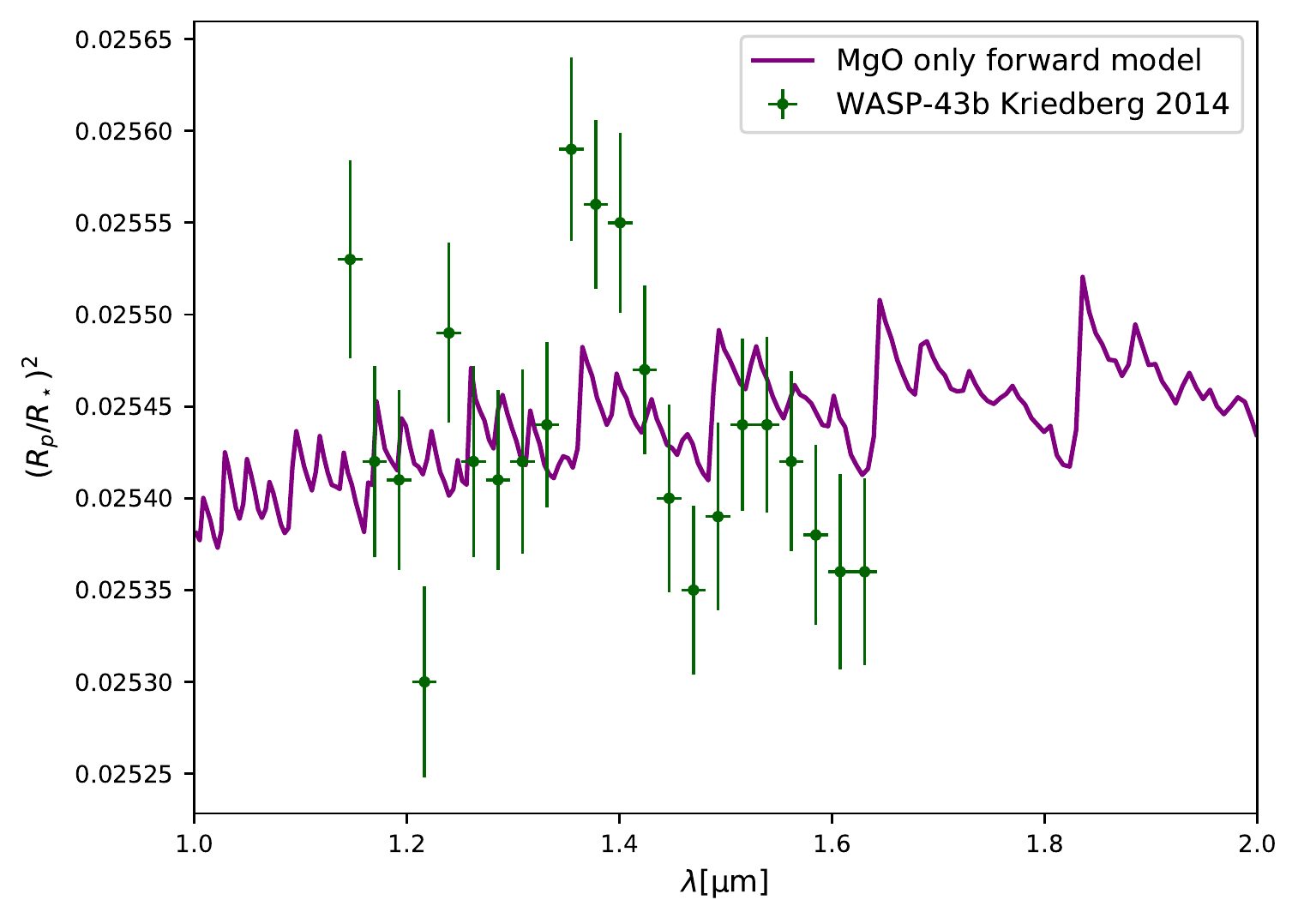}
        \includegraphics[width=0.5\textwidth]{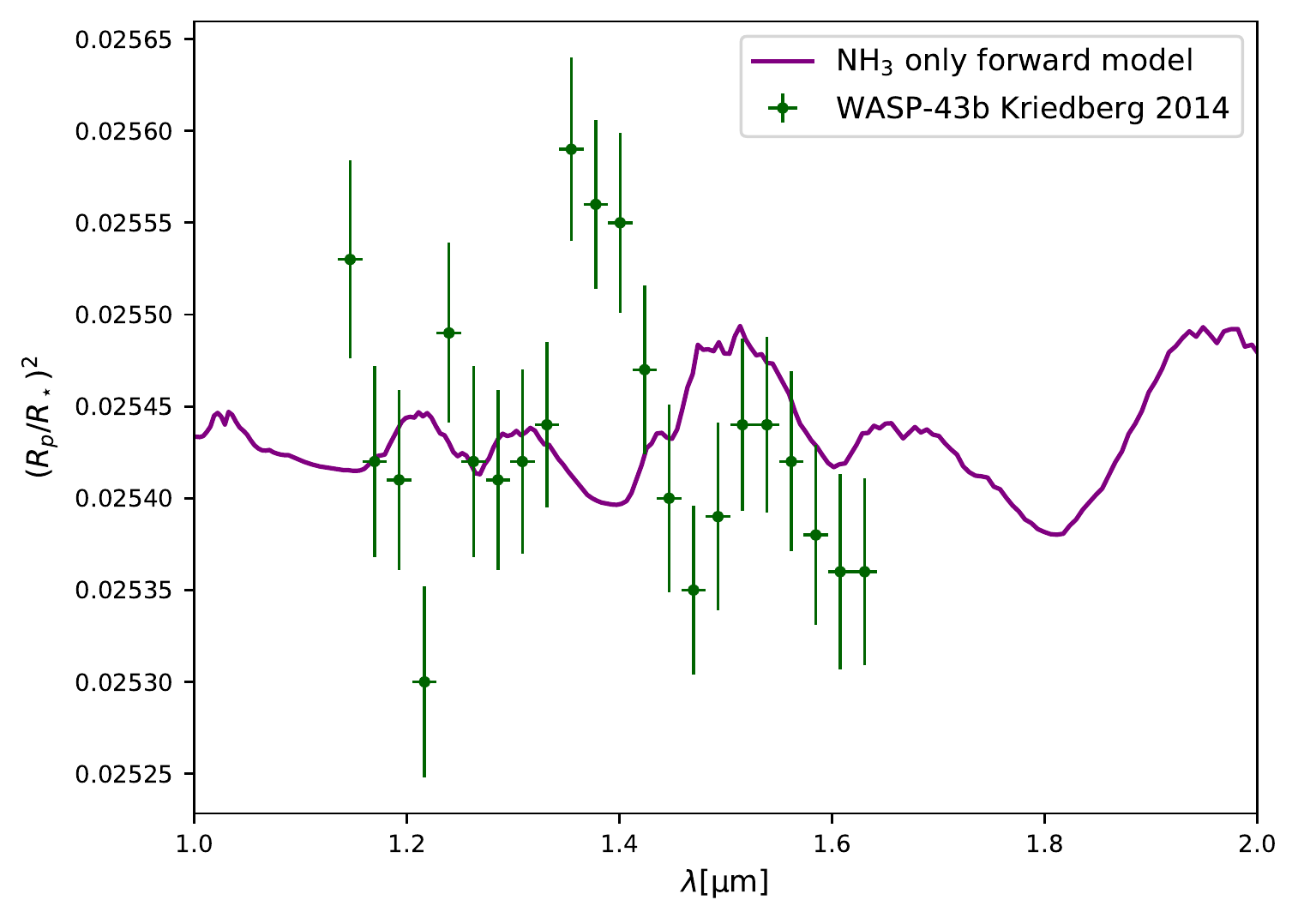}
        \includegraphics[width=0.5\textwidth]{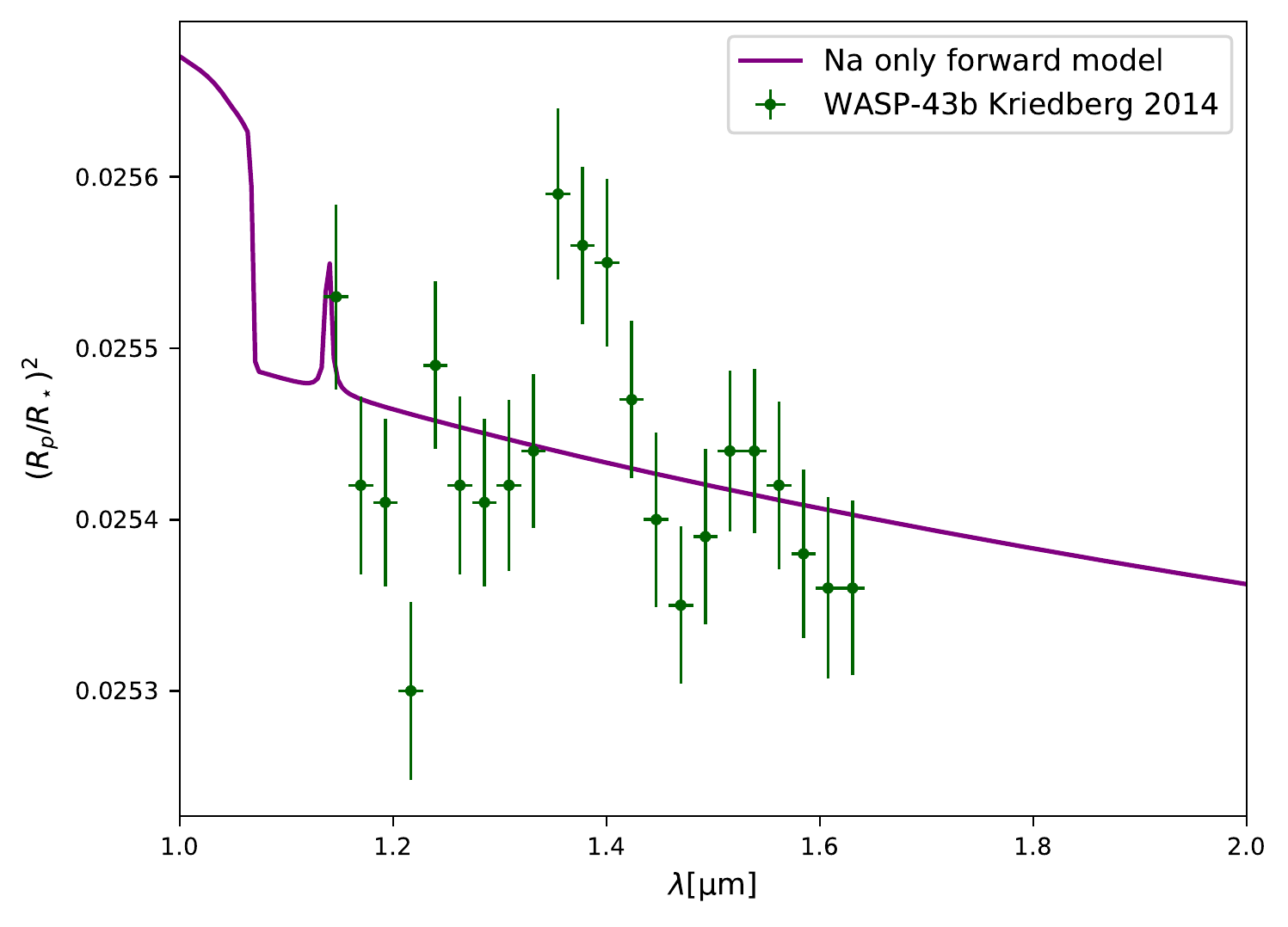}
        \includegraphics[width=0.5\textwidth]{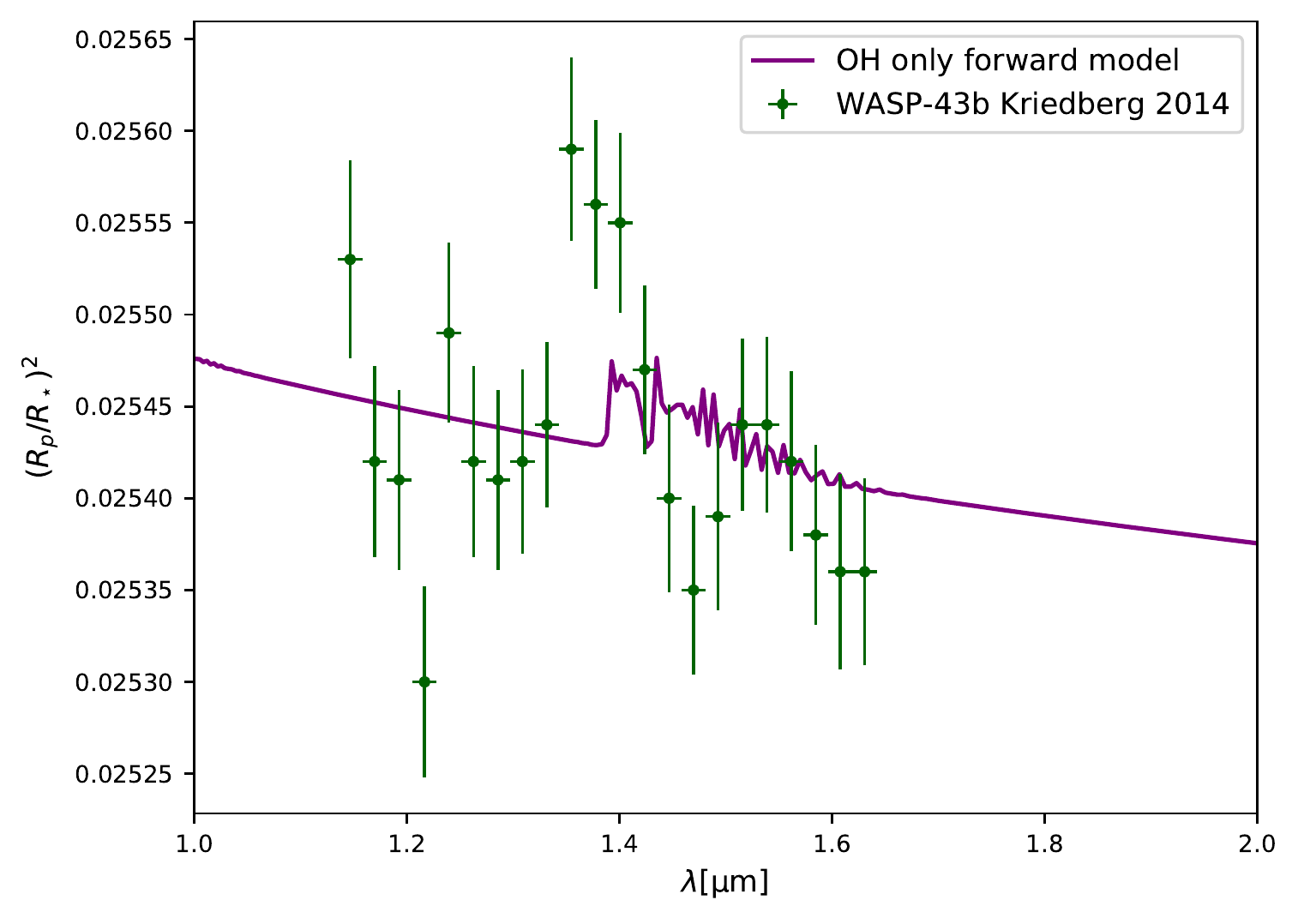}
        \caption{(continued) ARCiS forward models, each including one individual species, plotted alongside the transmission data for WASP-43b from \cite{14KrBeDe.wasp43b} in order to help assess which molecules to include in subsequent retrievals.}
        \label{fig:ARCiS_assess}
\end{figure}

\begin{figure}\ContinuedFloat
        \includegraphics[width=0.5\textwidth]{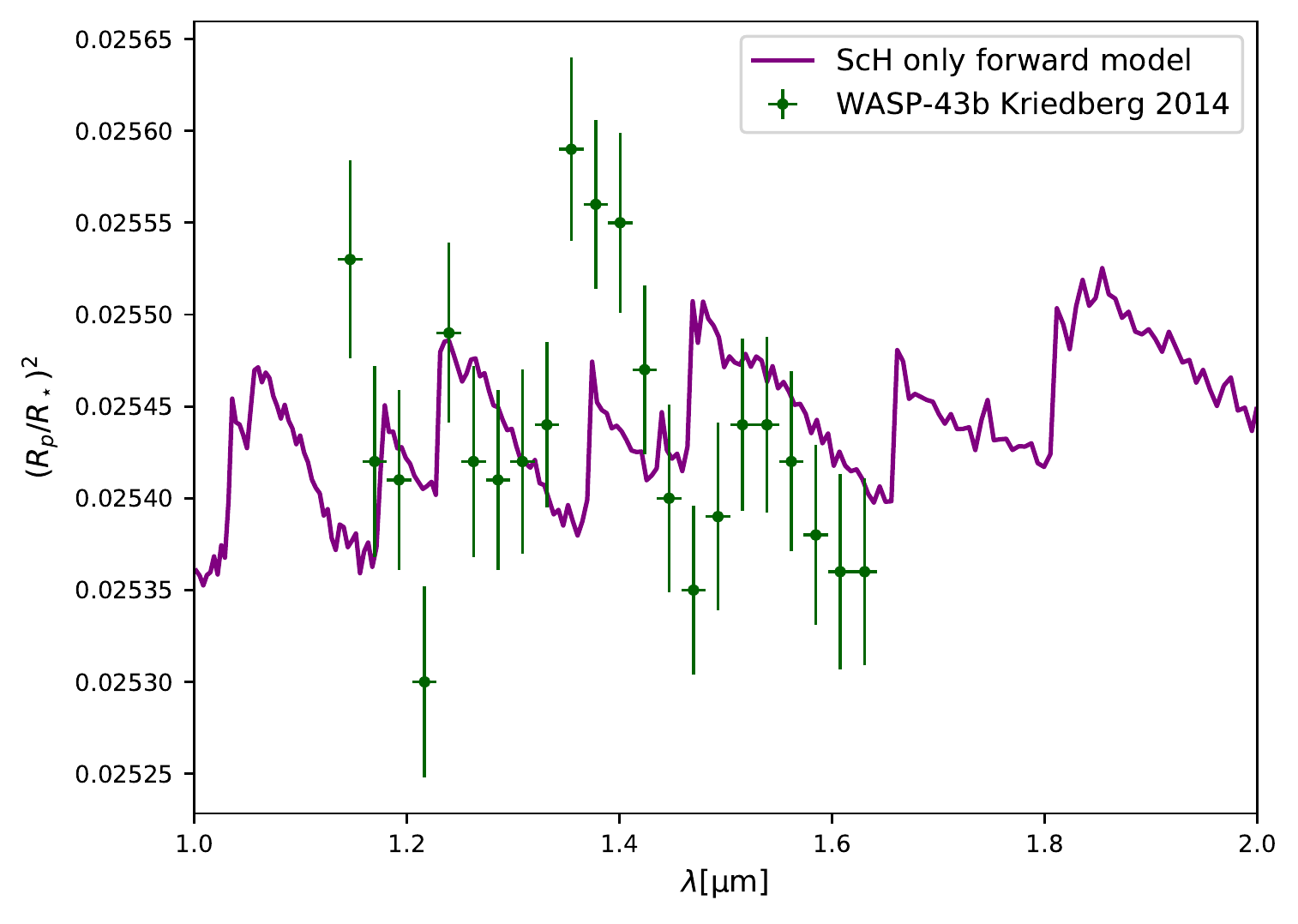}
        \includegraphics[width=0.5\textwidth]{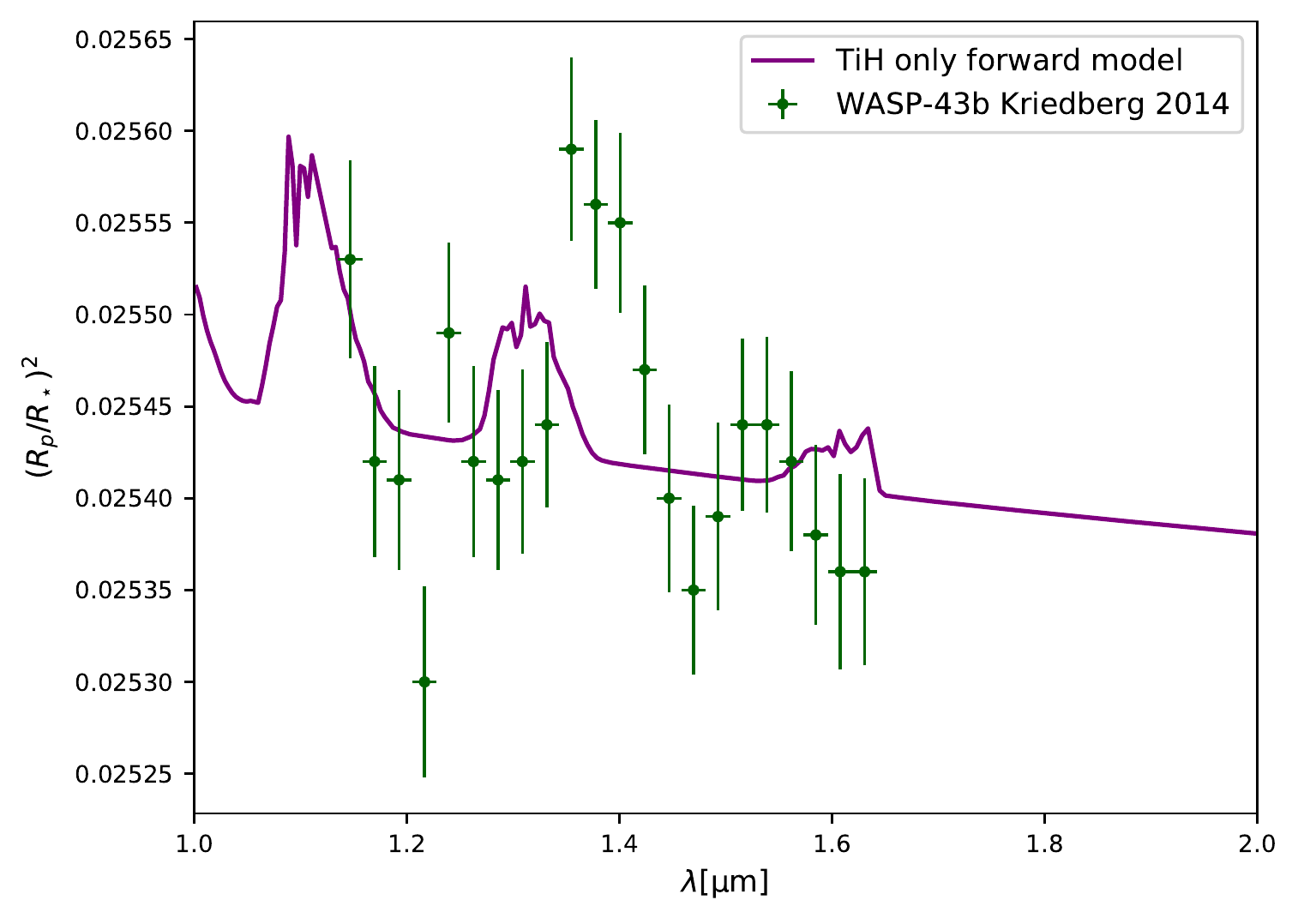}
        \includegraphics[width=0.5\textwidth]{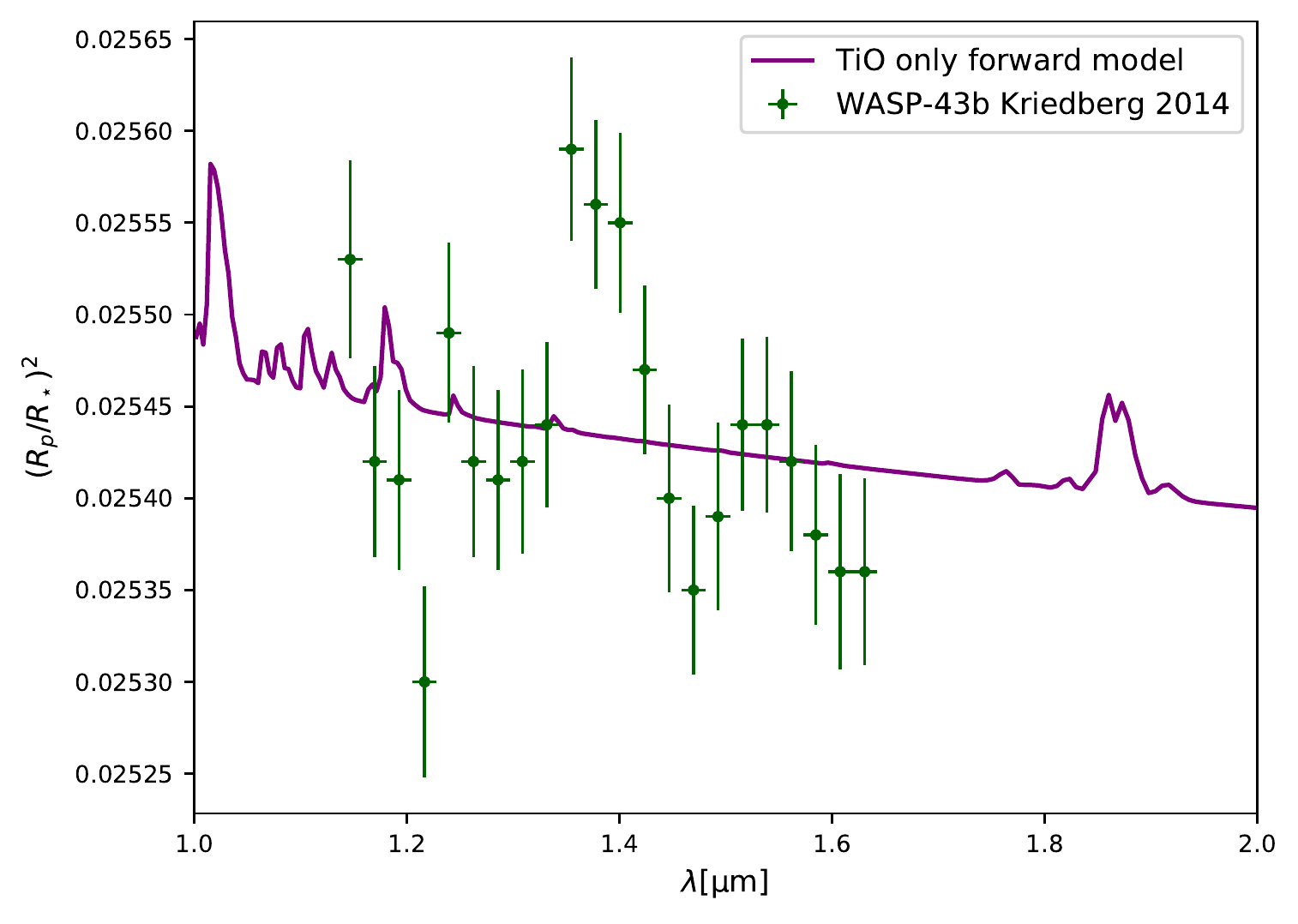}
        \includegraphics[width=0.5\textwidth]{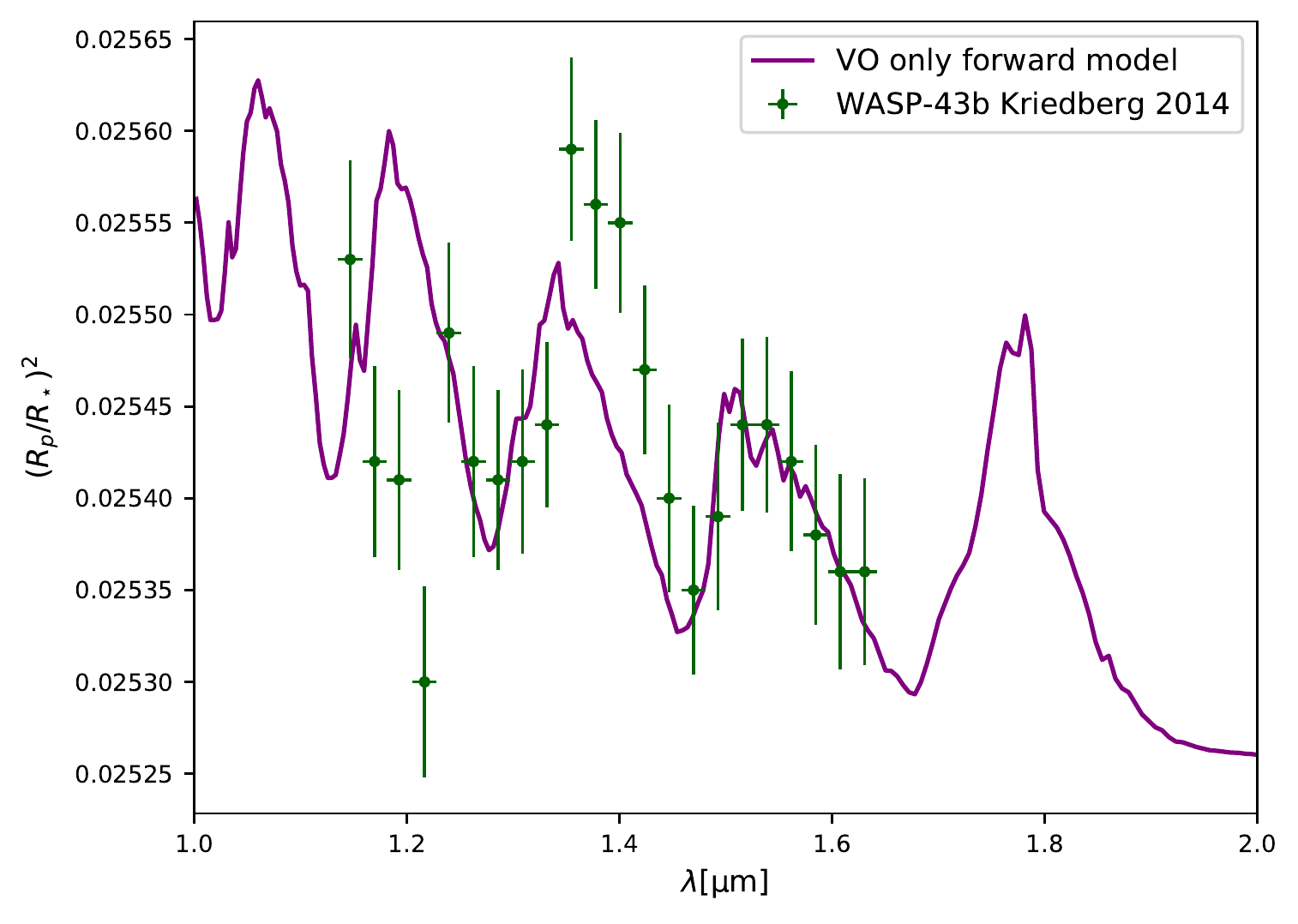}
        \caption{(continued) ARCiS forward models, each including one individual species, plotted alongside the transmission data for WASP-43b from \cite{14KrBeDe.wasp43b} in order to help assess which molecules to include in subsequent retrievals. }
        \label{fig:ARCiS_assess}
\end{figure}

\end{appendix}

\end{document}